\documentclass[12pt,preprint]{aastex}

\newcommand       \mic        	 {$\mu$m}
\usepackage{geometry}                
\usepackage{graphicx}
\usepackage{amssymb}
\usepackage{epstopdf}
\usepackage{epsfig}
\usepackage{amsmath}
\usepackage{graphicx}

\newcommand       \Msun        	{$M_{\odot}$}

\shorttitle{IR Observations of SNR~1987A (II)}
\shortauthors{Bouchet et al.}

\received{ }
\begin{document}

\title{SN~1987A AFTER 18 YEARS: MID-INFRARED {\it GEMINI} and 
{\it SPITZER} OBSERVATIONS OF THE REMNANT 
\footnote{Based on observations obtained at the {\it Gemini} 
Observatory, which is operated by the Association of Universities for
Research in Astronomy (AURA), Inc. under cooperative agreement with
the NSF on behalf of the {\it Gemini} partnership: the National Science
Foundation (United States), the Particle Physics and Astronomy
research Council (United Kingdom), the National Research Council
(Canada), CONICYT (Chile), the Australian Research Council
(Australia), CNPq (Brazil), and CONICET (Argentina).}}

\author{Patrice Bouchet\altaffilmark{1,2}, Eli Dwek\altaffilmark{3}, John Danziger\altaffilmark{4}, Richard G. Arendt \altaffilmark{5}, I.  James M.~De
Buizer\altaffilmark{6},  Sangwook Park\altaffilmark{7},  Nicholas B. Suntzeff\altaffilmark{2}, Robert P. Kirshner\altaffilmark{8}, and Peter Challis \altaffilmark{8}}

\altaffiltext{1}{GEPI, Observatoire de Paris, Site de Meudon, 5 Place Jules Janssen,
F-92195, Meudon, France; Patrice.Bouchet@obspm.fr}
\altaffiltext{2}{Cerro Tololo Inter-American Observatory (CTIO),
National Optical Astronomy Observatory (NOAO), Casilla 603, La Serena,
Chile; CTIO is operated by the Association of
Universities for Research in Astronomy (AURA), Inc. under cooperative
agreement with the National Science Foundation.}
\altaffiltext{3}{Observational Cosmology lab., Code 665; NASA Goddard
Space Flight Center, Greenbelt, MD 20771, U.S.A.}
\altaffiltext{4}{Osservatorio Astronomico di Trieste, Via Tiepolo, 11,
Trieste, Italy} 
\altaffiltext{5}{Science Systems \& Applications, Inc. (SSAI), Code 665, NASA Goddard
Space Flight Center, Greenbelt MD, 20771, U.S.A.}  
\altaffiltext{6}{Gemini Observatory, Southern
Operations Center, c/o AURA, Casilla 603, La Serena, Chile}
\altaffiltext{7}{Department of Astronomy and Astrophysics, Pennsylvania State University, 525 Davey
Laboratory, University Park, PA 16802, USA} 
\altaffiltext{8}{Harvard-Smithsonian, CfA, 60 Garden St., MS-19, Cambridge, MA 02138, U.S.A.}

\begin{abstract}
    
Using the Gemini South 8m telescope, we obtained high resolution 11.7
and 18.3~\mic\ mid-IR images of SN 1987A on day 6526 since the explosion.
All the emission arises from the equatorial ring. Nearly contemporaneous
spectra obtained at 5--38~\mic\ with the {\sl Spitzer} Space Telescope show  
that
this is thermal emission from silicate dust that condensed out in the
red giant wind of the progenitor star. The dust temperature is $166^{+18}_{-12}$~K, 
and the emitting dust mass is $(2.6^{+2.0}_{-1.4}) \times 10^{-6}$ \Msun.   
Comparison of
the Gemini 11.7~\mic\ image with Chandra X-ray images, Hubble UV- optical
images, and ATCA radio synchrotron images shows generally good
correlation across all wavelengths. If the dust resides in the diffuse
X-ray emitting gas then it is collisionally heated. The IR emission
can then be used to derive the plasma
temperature and density, which were found to be in good
agreement with those inferred from the X-rays.  Alternatively, the dust
could reside in the dense UV-optical knots and be heated by the  
radiative
shocks that are propagating through the knots. In either case the
dust-to-gas mass ratio in the CSM around the supernova is significantly
lower than that in the general interstellar medium of the LMC,
suggesting either a low condensation efficiency in the wind of the
progenitor star, or the efficient destruction of the dust by the SN
blast wave. Overall, we are witnessing the interaction of the SN blast
wave with its surrounding medium, creating an environment that is
rapidly evolving at all wavelengths.

\end{abstract}


\section{INTRODUCTION}
Since its explosion, SN~1987A has evolved from a supernova (SN) dominated by the emission from the radioactive decay of 
$^{56}$Co, $^{57}$Co and $^{44}$Ti  in the ejecta to a supernova remnant whose emission is dominated by the interaction of the supernova blast wave with its surrounding medium. The medium surrounding the SN is dominated by 
the well-known 
``circumstellar envelope" (CSE), which consists of an inner equatorial ring (ER) flanked
by two outer rings \citep{Bur95}, possibly part of an hour-glass structure.

The collision between the ejecta of SN~1987A and the ER 
predicted to occur sometime in the interval 1995-2007  
\citep{Gae97,Bor97} is now underway. At UV-optical (UVO) wavelengths, ``hot spots" have appeared inside
the ER \citep{Pun97}, and their brightness varies on time scales of a
few months \citep{Law00}.  
New hot spots continue to appear as
the whole inner rim of the ER lights up. The visible-light {\it HST} image obtained in 2004 Feb 20
reveals a necklace of such hot spots which nearly fill a lighted ring.
Ongoing monitoring at X-rays with the {\it Chandra} and at radio frequencies, shows
that the evolution of the emission from the ER follows a similar pattern at all wavelengths. 

There exist very few mid-infrared (IR) observations of supernovae in general.
Therefore SN~1987A, the closest known supernova in 400 years, gives us an
opportunity to explore the mid-IR properties of supernovae and  
the dust in their ejecta and surrounding medium
with the help of the newest generation of large-aperture telescopes and
sensitive mid-IR instrumentation like the T-ReCS on the {\it Gemini},  in combination with IR data obtained from
 {\it Spitzer Space Telescope}({\it SST}) \citep{Wer04}.
The T-ReCS observations of the mid-IR emission from SNR~1987A are part of our continuous 
 monitoring of the SN and its surrounding medium.
The first detection and analysis of mid-IR emission at the position of the 
supernova has been reported in \citep{Bou04} (hereafter Paper I).

The origin of the mid-IR emission could be line emission from
atomic species, synchrotron or free-free continua, or thermal
emission from dust which is probably the dominant source of emission.   
In general, there are several scenarios for the origin and the heating mechanism of the dust giving rise to the late time mid-IR emission in Type II supernovae \citep{Gra86,Ger00}. Thermal mid-IR emission could be: (1) the emission from SN-condensed dust that is collisionally heated by reverse shocks traveling through the SN ejecta;  (2) the emission from circumstellar/interstellar dust heated by the interaction of the expanding SN blast wave with the ambient medium; and (3) the delayed emission (echo) from circumstellar dust  radiatively heated by the early UVO supernova light curve.

Our imaging observations provide strong constraints on these possible scenarios for the mid-IR emission. They show that the bulk of the mid-IR emission is not concentrated on the center of the explosion, but arises from a ring around the SN. The morphology of the mid-IR emission can therefore be used to eliminate several scenarios for its origin. Combined with {\it Spitzer} spectroscopy, our observations can be used to determine the dust composition and temperature distribution in the ring. The  {\it Chandra} X-ray, and {\it HST} UVO data provide important constraints on the physical conditions of the medium overtaken by the supernova blast wave. These constraints can be used to determine the physical association and the heating mechanism of the dust giving rise to the mid-IR emission. 

The paper is organized as follows: We first describe in \S2 the imaging observations of the SN obtained by T-ReCS, and compare its IR morphology to that at radio, X-ray, and UVO wavelengths. Lower resolution {\it Spitzer} mid-IR imaging observations detected SNR~1987A (including the ring) as an unresolved point source, and are used in conjunction with spectroscopic data to determine the possible contribution of lines, and the composition of the dust giving rise to the continuum emission. In \S3 we describe the procedure used for the analysis of the spectroscopic and imaging data, determining the dust composition, and presenting maps of dust temperature, IR opacities, and dust column densities. The IR image of the circumstellar medium around SN~1987A has a morphology similar to the {\it Chandra} X-ray and {\it HST} UVO images. The limited mid-IR resolution does not allow us to unambiguously  determine whether the dust resides in the X-ray emitting gas, or in the UV-optical line emitting knots in the ER. We therefore resort in \S4 to an analysis of possible dust heating mechanisms and to calculating the inferred dust masses and dust-to-gas mass ratios for several possible scenarios. In \S5 we discuss the evolution of the supernova and its environment as manifested from the observed light curves at various wavelengths.  The results of our paper are summarized in \S6.

\section{OBSERVATIONS}

\subsection{Mid-Infrared {\it Gemini} Observations}

The T-ReCS mid-IR imager/spectrometer at the {\it Gemini} 8m telescope
offers a combined telescope and instrument with diffraction limited
imaging ($\sim0.3"$ resolution) and superbly low thermal emissivity.  On
2003 Oct 4 (day 6067), we imaged SNR~1987A with T-ReCS as part of the
instrument's System Verification program and we reported on the detection
of 10 and 20~$\micron$ emission from the ER, and on a 10~$\micron$ emission
from the supernova's ejecta (Paper I). Subsequent observations
were carried out in January 6, 2005 (day 6526) in the narrow Si5 filter 
($\lambda_{eff}$ = 11.66 \mic; $>$ 50\% transmission at $\lambda$ = 11.09-12.22 \mic),
and in February 1, 2005 (day 6552) in the Qa filter ($\lambda_{eff}$ = 18.30 \mic;
$>$ 50\% transmission at $\lambda$ = 17.57-19.08 \mic). 
Results are presented in Figures 1-a and 1-b.
These images show several luminous ``hot'' spots distributed over
the ring.  The calibrated flux density integrated within an aperture
of 1.3 arcsec radius is $F_{\nu}$($11.7~\micron$) = 18.4 $\pm 1.2$~mJy in the Si5 filter, and 
$F_{\nu}$($18.3~\micron$) = 53.4 $\pm 9$ mJy in the Qa filter. No color
correction was applied and this would most likely increase the flux
density.  The standard star used for the calibration of the 11.7~\mic\ measurement 
was HD~29291, whose flux density was taken to be 6.78 Jy at 11.7 \mic.
We used $\alpha$ CMa with a flux density of 44.3 Jy at 18.30 \mic\ for the 
flux calibration of the 18.3~\mic\ observation. 

The black body colour temperature corresponding to the measured fluxes at these two wavelengths is $T = 185$~K and the 
luminosity is $L_{BB} = 3.74 \times 10^{36}$ erg s$^{-1}$. 
We use the \citet{Mat90} extinction law with $\tau_{18.3} = \tau_{11.7} / 1.35$ and 
$A_{18.3}/A_J = 0.083$ and $A_{11.7}/A_J = 0.098$ to compute the black body temperature
and the optical depth for each individual pixel, resulting in the maps shown in Fig 1-c and 1-d. Note that in order to have the
iterative algorithm converge we must assume reasonable values as a starting point. We stress that Fig 1-c and
1-d show the {\it color} temperature and optical depth maps, which are slightly different from the maps related to the physical dust as calculated in Section 3. They are shown only for illustrating the results obtained from our data fitted to the simplest modeling (eg. black body).

In Figure~\ref{fig2} we compare our $11.7~\micron$ new data with the one obtained
in the broad N band (10~\mic) on October 2003 at day 6067 (Paper I). This figure shows a clear brightening in the
South-West region of the ER, superimposed on a general brightening all over the ring. Given that the two images are taken with significantly different filters we investigated whether the difference could be due to the different spectral coverage: the spectra obtained with {\it Spitzer} which are discussed in next section do not show any feature that could be a source of the difference. Thus we conclude that we are observing a true brightness change.

Figure~\ref{fig3}(a,b) displays our images in both filters with the contours of the 0.3--8 keV X-ray image from the 
{\it Chandra} Observatory obtained nearly simultaneously at day 6529 (January 9--13, 2005) 
\citep{Par05b}. The correspondences between our $11.7~\micron$ and that obtained by the {\it Chandra} is very good, but less so at $18.3~\micron$ because of the lower signal-to-noise ratio in that image. A more detailed discussion of the relation between the IR and X-ray images of the ER will be presented in \S3 below. 

Figure~\ref{fig4}(a, b) shows the contours of the image obtained in the 12 mm band (16-26 GHz) at
day 6003 (July 31, 2003) with the Australian Telescope Compact Array (ATCA) at
the Australian National Telescope Facility (ATNF)  \citep{Man05} superimposed on our 11.7 and 18.3~\mic\ T-ReCS images, respectively. 
The correspondences between our $11.7~\micron$ image with the synchrotron radio emission is not as
remarkable as it is with the {\it Chandra} image. Nevertheless, the
$18.3~\micron$ bright spot in the East side looks better correlated with the radio lobe than
X-ray spots do. 
Furthermore, 
an image obtained in May 5, 2004 (day 6298) at the same frequencies is posted on the ATNF web page
and is reproduced here in Figure~\ref{fig4}(c,d). This later radio image shows better correlation with our 11.7~\mic\ image than the radio image obtained on day 6003, probably because it was taken closer to the epoch of the mid-IR observations. This demonstrates the importance of evolutionary effects on the morphology of the emission at all wavelengths.

There is good overall agreement in shape and size between our IR
image and images obtained from the X-ray to the radio. 
The mean radii and approximate surface brightness distribution
(brighter on the east side) of the ring are similar at all wavelengths, demonstrating
that the dust is co-extensive with the gas components. The origin
of that brightness asymmetry may be related to an asymmetric
distribution of the ejecta or the CSM \citep{Par02,Par04}, and/or to a time-dependence 
effect caused by the tilt of the ER as argued by \citet{Pan91}.  

The most likely source for mid-IR radiation is thermal emission from warm
dust (see discussion below). The X-ray radiation is thermal emission
from very hot gas (\citet{Par05a}, and previous references therein), whereas the optical emission arises from the dense knots in the ER that are overrun by slower shocks. The radio emission is 
likely to be synchrotron radiation from shock-accelerated electrons spiralling
in the remnant's magnetic field as stated by \citet{Dun03} and \citet {Man05}.
\citet{Par02,Par04} argue that, until 2000 December, hard ($E > 1.2$~keV) X-ray and radio 
emissions were
produced by fast shocks in the CS HII region while the optical and
soft ($E < 1.2$~keV) X-ray emissions came from slower shocks in the denser ER.
\citet{Par03} note that as of 2002 Dec. 31 (day 5791) correlations between
the X-ray and the optical/radio images are more complex than the above
simple picture, which is expected as the blast wave is reaching the main
body of the inner ring. 

\subsection{Spitzer Observations}

\subsubsection{Imaging Data}

Imaging of SNR~1987A was carried out with Spitzer Space Telescope's 
MIPS instrument at 24 \mic\ \citep{Rie04} and IRAC instrument at 3.6 -- 8 \mic\
\citep{Faz04}  
(AORIDs = 5031424 and 5030912).
Almost one year later SNR~1987A was again imaged with IRAC, 
serendipitously near the edge of the field of observations 
targeting other sources (AORIDs = 11191808
and 11526400).
All these data were obtained from the {\it Spitzer} data archive.
The corresponding images are shown in Figure~\ref{fig17}(a-f), which shows
also a near-IR image from the {\it Hubble} Space Telescope for comparison. 
At 24 and 8 \mic\ SNR~1987A was detected as an unresolved 
point source amid a field of complex cirrus emission.
At 5.8 \mic, the source appears very slightly distorted, and at 
4.5 and 3.6 \mic\ the SN appears to be swamped by the emission 
of companion stars 2 and 3. The flux densities at 24, 8, and 5.8 \mic\ 
were measured using SExtractor \citep{Ber96} to perform aperture photometry 
on the post-BCD images. 
Aperture radii used were 4, 5, and 6 pixels ($4.8''$, $6.0''$, and  
$14.7''$)
at 5.8, 8.0, and 24 $\mu$m respectively.
Approximate aperture 
corrections of 1.10, 1.07, and 1.14 were applied using information 
from the IRAC and MIPS Data Handbooks. 
The calibrated flux densities are given in Tab.~\ref{tab1}. 

\subsubsection{Spectroscopic Data}

Observations of SNR~1987A were also performed with {\it Spitzer}'s 
Infrared Spectrograph (IRS) in its 
short (wavelength) -- low (resolution), short-high, 
and long-high modes \citep{Hou04}. These data were also obtained from 
the {\it Spitzer} data archive (AORID = 5031168).
Figure~\ref{fig18} shows the slit positions for the different sets of observations. 
For each spectral mode we extracted 
SN~1987A spectra from the 2--D coadded post-BCD images
using SPICE (http://ssc.spitzer.caltech.edu/postbcd/spice.html). 
For the short--low data, the source is placed at 4 positions 
along the slit to generate spectra  at two positions for two 
different spectral orders. For each slit position, a 2--D background
image was generated from the median value of the images for the
other three slit positions. These backgrounds were subtracted prior
to extracting spectra with SPICE. For the high resolution observations,
only two slit positions are observed within a much narrower slit.
So for these observations, SPICE was used to extract the SN spectrum 
from the columns occupied by the unresolved source, and a 
background spectrum from columns toward the opposite end of the slit.
The background spectra were then subtracted from the source spectra.
For all modes the spectra from the two slit positions (per spectral order)
were combined using a weighted ($1/\sigma^2$) average, and (generally
noisy) data where spectral orders overlapped were discarded. For the
short-hi data, an empirically determined scaling factor of 1.38 was applied
before averaging to bring results from both slit positions into agreement. 
An additional scale factors of 1.46 was applied to the short--high spectrum to normalize 
it to the short--low data, and a subsequent factor of 1.25 was applied to 
long-high data to normalize to the short-high data. These latter scaling 
factors are to be expected if the subtracted background was partially 
contaminatead by the source, or in the case that the smaller high resolution
slits were slightly misaligned. Finally for the sake of comparison with the
broad band measurements and dust models, the high resolution spectra 
were median--binned using intervals of 12 and 11 wavelength samples  
for the short and long wavelengths respectively. 

Figure~\ref{fig5} shows the overall calibrated spectrum and Figure~\ref{fig6} displays individual profiles 
and identifications of the main emission lines detected with IRS from the {\it Spitzer} Space 
Telescope: it shows that the T-ReCS observations are dominated by the dust continuum 
emission, and not the lines. The [Ne II] 12.81 $\mu$m and [Ne III] 15.56 $\mu$m lines are however clearly 
seen. A weak [Si II] 34.81 $\mu$m line remains after background subtraction, while [S III] 
33.48 $\mu$m disappears entirely. Two strong lines are seen near 26 $\mu$m: the redder line 
is [Fe II] at 25.99 $\mu$m and the bluer line is [O IV] at 25.89 $\mu$m. 
Both lines could be arising from [Fe II] (or both [O IV]) if from fast 
moving ejecta on near and far sides of the explosion, but the lack of splitting of the other 
lines makes this seem unlikely. A weak third component seems present here as well. 
If fitted as an unresolved line this may be [F IV] at 25.83 $\mu$m. However, it may also
be fitted as the wing of a very broad line (FWHM $\sim$ 2800 km s$^{-1}$) underlying 
the narrower [O IV] and [Fe II] lines. Such high velocity would indicate an association
with the SN ejecta rather than the ER for this line.
Line fluxes, centers, and widths have been calculated by gaussian fits 
using SMART \citep{Hig04} and results are given in Tab.~\ref{tab2}.


\subsection{Near-IR CTIO Observations}

Near-IR J($1.25~\micron$), H($1.65~\micron$), and K($2.2~\micron$)
imaging observations of SNR~1987A were obtained on 2005, January 3--5
with ISPI attached to the Blanco 4-m telescope at the {\it Cerro Tololo Interamericain Observatory}. Results are displayed in Figure~\ref{fig7}. We clearly see the ER at J and K, while
most of the emission detected in the H band arises from the supernova itself.  
These images have been deconvolved by using the ``Multiscale Maximum Entropy Method" algorithm developed by \citet{Pan96}.
The flux densities measured in the CTIO data are ($F_J = 0.33 \pm 0.25$ mJy; $F_H = 0.11 \pm 0.05$ mJy;
$F_K = 0.51 \pm 0.2$ mJy). These fluxes refer to the total of the
ER + ejecta in each band (although each band seems dominated by one or the other).
Images obtained with HST on November 11, 2005 through the filters
F110W, F160W, and F205W, are also shown in this figure, and show that
compared to the ring, the ejecta are relatively brighter in H band (F160W) than in other bands. 
This clearly demonstrates the validity and the
power of the deconvolution algorithm applied to our CTIO data.
Note, however, that the HST bands are not exactly J, H, K: one significant
difference is that the Paschen $\alpha$ line is missed between the
ground based H and K bands, but would be seen in HST's F205W band.

The Spitzer mid-IR spectrum indicates that dust in SN 1987A
and the ER is too cool (see Section 3) to emit significantly in these
near-IR bands. We believe that line emission dominates these
broadband near-IR observations, and that differences in the
composition, temperature, and density of the ejecta vs. CSM
lead to the different relative brightnesses of these structures.
The young supernova remnants Cas A and Kepler may be the
closest analogs we have for interpreting SN 1987A's near-IR emission.
Near-IR spectra and imaging of these SNRs by \citet{Ger01} and \citet{Rho03}
reveal similar variation in near-IR
colors. Circumstellar material (e.g. Cas A's quasi-stationary flocculi)
is bright in J due to He I 1.083 $\mu$m, and emits strong [Fe II] lines
in both J and H bands. Weaker hydrogen lines can be detected in all
bands.  The ejecta in several of Cas A's
fast-moving knots emit strong [S II] lines at 1.03 $\mu$m in J,
and weaker lines in [Fe II] and Si in various ionization states.
Thus, the near-IR emission of the circumstellar ER may be dominated
by H and He I lines in J and H with synchrotron emission possibly
augmenting the K band \citep{Ger01,Rho03}.  
The synchrotron processes deserve consideration in SN~1987A particularly 
in view of the fact shown in Figure~\ref{fig4} that the radio morphology 
of SN~1987A reproduces the ring shape now.
Also, the H2 molecule has many emission lines in the K-band and longer wavelengths. 
Spectra are clearly needed for
testing the plausibility of the contribution from this molecule or otherwise. 
The first overtone band head of molecular CO lies at 2.29$\mu$m and may be 
also contributing to K-band emission in the ring. 
The relatively strong H band emission of SN 1987A's ejecta
could reflect [Fe II] emission, because a large fraction of the ejecta
should be Fe from the core of the explosion, whereas Cas A's FMKs
generally consist of lighter metals originating
in the outer layers of the progenitor.


\section{DATA ANALYSIS}

\subsection{The Dust Properties}

Figure 5 clearly demonstrates that the mid-IR emission is dominated by thermal emission from dust. 
The specific luminosity of a single dust particle of radius $a$ at temperature $T_d$, at wavelength $\lambda$ is given by:
\begin{eqnarray}
{\ell}_{\nu}(\lambda) & = & 4 \pi a^2 \pi B_{\nu}(\lambda,\ T_d) Q(\lambda) \\ \nonumber
 & = & 4 m_d \kappa(\lambda,\ a) \pi B_{\nu}(\lambda,\ T_d)
\end{eqnarray}
where $B_{\nu}(\lambda,\ T_d)$ is the Planck function, $Q(\lambda)$ the dust emissivity at wavelength $\lambda$, and  $\kappa(\lambda,\ a) \equiv 3Q(\lambda)/4 \rho a$ is the dust mass absorption coefficient, where $\rho$ is the mass density of the dust particle. In the Rayleigh limit, when $a < \lambda$, $\kappa$ is independent of particle radius. Figure~\ref{fig8} illustrates the $\kappa (\lambda)$ curves for amorphous
carbon \citep{Rou91}, graphite and silicate grains \citep{Lao93} over the 5 to 30~$\micron$ wavelength region. The figure shows the distinct optical properties of these dust particles over the mid-IR wavelength regime which can greatly facilitate the identification of the emitting material with even limited broad band filters.

For an optically thin point source, the flux density, $F_{\nu}(\lambda)$, at wavelength $\lambda$ is given by:
\begin{equation}
F_{\nu}(\lambda) = 4\ M_d\  { \kappa(\lambda)\ \pi B_{\nu}(\lambda,\ T_d)\over 4 \pi D^2}
\end{equation}

\noindent
where $M_d$ is the dust mass, and $D$ is the distance to the supernova, taken to be $D =51.4$ kpc \citep{Pan99}. 

For an extended optically thin source with an angular size $\Omega$, the surface brightness, $I_{\nu}(\lambda)$, is given by:
\begin{equation}
I_{\nu}(\lambda) = \Omega\ \tau_d(\lambda) B_{\nu}(\lambda,\ T_d)
\end{equation}
where $\tau_d(\lambda)$, the dust optical depth, is given by:
\begin{equation}
\tau_d(\lambda) = {M_d\ \kappa(\lambda)\over \Omega \ D^2}
\end{equation}

\subsection{Spectral Analysis}

We fitted the integrated T-ReCS flux densities with a population of dust particles consisting of a single population of bare graphite, silicate or amorphous carbon grains, using equation (2). Optical properties of silicate and graphite grains were taken from
\citet{Lao93}, and those for amorphous carbon (BE) were taken from \citet{Rou91}.
The results are given in Figure~\ref{fig9}, which shows that the T-ReCS observations alone cannot discriminate between the different dust compositions. Comparison with the {\it Spitzer} observations clearly demonstrates that the IRS observations can be  
well fit with a silicate dust composition (Figure~\ref{fig10}), ruling out  graphite or carbon dust as major dust constituents in the CSM. 

Figure~\ref{fig10} shows the fit of the silicate dust spectrum to the {\it Spitzer } IRS spectrum.  The silicate temperature is $T_d = 180^{+20}_{-15}$~K, and the dust mass is $M_d = (1.1^{+0.8}_{-0.5})\times 10^{-6} M_\odot$. The global parameters resulting from our model for silicates are given in Table~\ref{tab3}. Figure~\ref{fig11} shows the residuals of the fit, obtained by subtracting the silicate model fit from the data.  The residuals are small, and their spectrum is too sharply peaked at short wavelengths 
to be fitted with any blackbody. The residuals are also broader than typical atomic line widths, suggesting that they may be due to solid state features in the dust, reflecting differences in  the crystalline structure of the silicates in the CSM from the average interstellar silicate dust used in the model. This figure shows also that the residuals cannot be fit by emission from amorphous carbon dust (shown by the green line in this figure).

\subsection{Image Analysis}

Our spectral analysis showed that the mid-IR spectrum of the ER is dominated by silicate emission. We therefore used the 11.7  and 18.3~\mic\ images of the remnant to construct temperature, and dust opacity maps of the circumstellar ring using Eq. 3 to derive the dust temperature, and Eq. 4 to derive the dust optical depth. In calculating these quantities, we applied a background threshold of 0.03~mJy/pix at 11.7~\mic\ and of 0.08~mJy/pix at 18.3~\mic. The mass of the ring was calculated from the optical depth, over a surface area of 269 pixels, corresponding to the number of pixels that had a flux exceeding the respective thresholds at each wavelength. The average dust temperature in these maps is $T_d = 166_{-12}^{+18}$~K. The average 11.7~\mic\ optical depth per pixel is $\tau_d = (5.5^{+4.2}_{-2.7})\times10^{-6}$, and the total dust luminosity is $L_d = (2.3^{+0.5}_{-0.4}) \times 10^{36}$~erg~s$^{-1}$, giving a dust mass $M_d = (2.6_{-1.4}^{+2.0}) \times10^{-6}$~\Msun, in good agreement with the total mass obtained from the spectral analysis of the ER. 
Figure~\ref{fig12} shows the maps of the silicate dust temperature (a) and optical depth (b) in  the ER.

\section{THE ORIGIN OF THE MID-INFRARED EMISSION}

The data presented in this paper show unambiguously that the emission is thermal emission from dust. At issue are the location and heating mechanism of the dust. The forward expanding non-radiative blast wave is currently interacting with the circumstellar material and the knots in the ER. The interaction of the blast wave with the knots transmits lower velocity radiative shocks into these dense regions, producing soft X-rays and the ``hot spots" seen in the {\it HST} images. The interaction of the blast wave with the dense knots also generates reflected shocks that propagate back into the medium that was previously shocked by the expanding SN blast wave. The complex morphology and density structure of the ER gives rise to a multitude of shocks characterized by different velocity, temperatures, and post shock densities. 

The mid-IR images cannot determine the location of the radiating dust, whether it resides in the X-ray emitting gas or in the denser UVO emitting knots. Therefore, we can not, a priori, assume a particular dust heating mechanism:  collisional heating in the shocked gas, or radiative heating in the radiative shocks. 

\subsection{Dust Heating Mechanism}

The relative importance of the  two dust heating mechanisms is given by the ratio \citep{Are99}:
\begin{equation}
{\cal R} \equiv {H_{rad}\over H_{coll}} = {n_e\ n_H\ \Lambda(T_e)\ P_{abs} \over n_d\ n_e\ \Lambda_d(T_e)} 
\end{equation}

\noindent
where $H_{rad}$ and $H_{coll}$ are, respectively, the radiative and collisional heating rates of the dust, $n_e$, $T_e$, and $\Lambda(T_e)$ are, respectively, the electron density, temperature, and the atomic cooling function (erg cm$^3$ s$^{-1}$) of the gas, $P_{abs}$ is the fraction of the cooling radiation that is absorbed by the dust,  and $\Lambda_d(T_e)$ is the cooling function of the gas via electronic collisions with  the dust and given by:
\begin{equation}
\Lambda_d(T_e) = 2 {\bar v_e}\pi a^2 kT_e \langle h\rangle
\end{equation}
where $n_d$ the number density of dust grains, $a$ their average radius, ${\bar v_e} = ({8kT_e/\pi m_e})^{1/2}$ is the mean thermal speed of the electrons, and $\langle h \rangle \lesssim 1 $ is the collisional heating efficiency of the dust which measures the fractional energy of the electrons that is deposited in the dust \citep{Dwe87a}.

 In the following we will examine possible locations of the dust giving rise to the IR emission. For each possible site, the X-ray emitting gas or the optical knot, we determine the dominant cooling mechanism, the temperature of the dust, and the inferred dust mass. A site is viable if it can maintain a dust temperature between $\sim$ 150 and 200~K, the range of values reflecting the range of dust temperatures derived from the T-ReCS and {\it Spitzer} observations, and if the derived dust mass does not violate any reasonable abundances constraints.
 
\subsection{Dust in the X-ray Emitting Gas}

The morphological similarities between the $11.7~\micron$ emission,  dust temperature, and optical depth maps  on one hand and the X-ray maps of the supernova on the other hand suggests that the dust giving rise to the IR emission may be well mixed with the X-ray emitting gas. 

For an optically thin plasma $P_{abs} \approx \tau_d =  n_d \pi a^2\langle Q \rangle \ell $, where $\langle Q \rangle$ is the radiative absorption efficiency of the dust averaged over grain sizes and the radiation spectrum, and $\ell$ is a typical dimension of the emitting region. Inserting this value in eq. (5) we get:
\begin{eqnarray}
{\cal R} & = &  {n_e \Lambda(T_e) \ell \over 2 v_e kT_e}\ {\langle Q \rangle\over \langle h \rangle} \\ \nonumber
 & = &  {\sqrt{\pi m_e\over 32}}\ \times {n_e \Lambda(T_e)\ \ell \over (kT_e)^{3/2}}\ \times {\langle Q \rangle\over \langle h \rangle}
\end{eqnarray}

For average conditions in the X-ray plasma (see below), characterized by plasma temperatures and densities of $T_e \approx 10^7$~K, and $n_e \approx 300-10^3$~cm$^{-3}$, we get $\Lambda(T_e) \approx 4\times 10^{-23}$ erg~cm$^3$~s$^{-1}$, and $\langle h \rangle \gtrsim$ 0.1 for dust particles with radii larger than 0.05~\mic. Adopting a value of $\langle Q \rangle \approx $~1, gives an upper limit of 
\begin{equation}
{\cal R} \lesssim 2.1\times10^{-20} \ \ell
\end{equation}
The size of the  X-ray emitting  plasma is less than the radius of the ER which is $\ell \lesssim$ 0.7~lyr = $7\times10^{17}$~cm, giving ${\cal R} \lesssim  0.14$, that is, the dust heating in the X-ray gas is dominated by electronic collisions.

\subsubsection{The Dust Temperature}

Figure~\ref{fig13} depicts contours of the temperature of collisionally heated silicate grains of radius $a$ = 0.10~\mic\ as a function  of plasma density and temperature. To calculate the energy deposited in the dust we used the electron ranges of 
\citet{Isk83} for energies between 20 eV and 10 keV, and those of \citet{Tab72} for higher incident electron energies. We also assumed that all incident electrons penetrate the dust (no reflection). Above gas temperatures of about $3\times 10^6$~K, the temperature of collisionally heated dust is independent of grain radius, and very well represented by a single dust temperature, since both the radiative cooling and the collisional heating rates are proportional to the mass of the radiating dust particle. Furthermore, at these plasma temperatures the dust temperature is essentially independent of gas temperature as well, and therefore an excellent diagnostic of plasma densities. 
The figure shows that for plasma temperatures above $\sim 3 \times 10^6$~K, dust temperatures between 150 and 200~K require electron densities of about 300 to 1400~cm$^{-3}$.   

X-ray observations show the presence of two X-ray emission components \citep{Par06}:  one associated with the ``slow shock" with an electron temperature 
of $\sim0.23$~keV and a density of $n_e \sim$ 6000~cm$^{-3}$, and the second, associated with the ``fast shock" with electron temperatures and densities of  $kT \sim2.2$~keV 
and $n_e \sim$ 280~cm$^{-3}$, respectively. 
The lastest {\it Chandra} data indicate \citep{Par06} that 
the ``fast shock" and the ``slow shock" of the model are becoming less distinguishable, as the
overall shock front is now entering the main body of the inner ring, that is, the electron
temperature of the soft component is increasing and that of the hard component is decreasing.
Albeit rather speculative, the overall temperature might thus be ``merging" onto an
``average"  temperature of $ kT \sim1.5$~keV $\sim 1.8\times 10^7$~K, \citet{Par06}, and intermediate electron densities. The range of densities expected for this ``average" shock is in very good agreement with that implied from the IR observations.   

\subsubsection{The Infrared-to-X-ray Flux Ratio (IRX)}

An important diagnostic of a dusty plasma is the infrared-to-X-ray ($IRX$) flux ratio \citep{Dwe87b}. For a given dust-to-gas mass ratio and grain size distribution, the $IRX$ ratio is defined as the infrared cooling to X-ray cooling in the 0.2--4.0 keV band, and is given by:
\begin{eqnarray}
IRX(T_e) & \equiv & {n_e\ n_d\ \Lambda_d(T_e)\over n_e\ n_H\ \Lambda_x(T_e)} \\ \nonumber
 & & \\ \nonumber
 & = & {\mu m_H Z_d\over \langle m_d \rangle}\ {\Lambda_d(T)\over \Lambda_x(T_e)} 
\end{eqnarray}
were $Z_d \equiv n_d \langle m_d \rangle/\mu n_H m_H$ is the dust-to-gas mass ratio, $\mu$ is the mean atomic weight of the gas, and $\langle m_d \rangle$ the mass of a dust particle, averaged over the grain size distribution. For a given value of $Z_d$, the $IRX$ ratio is only a function of plasma temperature, and ranges from a value of about 10 for $T_e = 10^6$~K to a value of $\sim 400$ for $T_e = 10^8$~K.  For young supernova remnants, \citet{Dwe87b} found that the IRX ratio is significantly larger than unity for 7 of
the 9 remnants considered in their paper, the other two having only an upper
limit on their IR emission.

The observed $IRX$ ratio can be obtained from the X-ray and IR fluxes from the SN. From the January 2005 {\it Chandra} data, we estimate the X-ray flux in the 0.2--4.0 keV X-ray band to be
 $7.24 \times 10^{-12}$ erg cm$^{-2}$ s$^{-1}$ 
after correcting for interstellar absorption by an H-column density of $N_H = 2.35 \times 10^{21}$ cm$^{-2}$. 

Note that these values are based on the two-temperature model as used in \citet{Par04}.
Fractional contributions from the soft ($kT_e$=0.3 keV) component in the
total flux is $\sim$70\% for the unabsorbed flux
(and thus for $L_X$). Thus, a contribution from each component in the 0.2--4 keV
X-ray flux seems to be significant rather than being dominated by one
of them. 

The total IR flux is $F_{IR} = 7.7 \times 10^{-12}$ erg~cm$^{-2}$~s$^{-1}$
which leads to an $IRX$ ratio $\sim$ 1. This value is
lower than the values reported in Paper I, in which $IRX$ = 6 for the decelerated slow shock
component, and $IRX$ = 3 for the blast wave shock in the
two-temperature model. It is much lower than the theoretical value of $\sim 10^2$, expected for a $T_e \approx 1.8\times10^7$~K plasma, which is observed in the young remnants Tycho and Cas~A. Other (mostly older) SNRs with measurable IR emission show 
somewhat lower $IRX$ ratios, but not as low as SNR~1987A.

Several effects could be the cause for this very low value of the $IRX$ ratio in SNR~1987A.
First, in remnants the $IRX$ ratio was calculated by \citet{Dwe87a} for an interstellar dust-to-gas mass ratio of 0.0075. The SNR~1987A blast wave is expanding into the circumstellar shell of its progenitor star, which, a priori, is not expected to have an interstellar dust-to-gas mass ratio. Moreover, when estimating the depletion of elements onto dust in the ER, we should compare the
expected dust abundances in the ER with the maximum available abundances for the LMC. General LMC abundances exist for B stars \citep{Rol96} and ISM \citep{Wel99} and, although controversy still remains \citep{Kor02}, the LMC metallicity is usually assumed to be
$0.5 - 0.7$ solar. 
Assuming that the fraction of metals locked up in LMC dust is the same as in the local ISM, and that the ER has the same metallicity as the LMC, then the IRX ratio in the ER should be about $0.5 - 0.7$ times that expected from Supernova Remnants in the Milky Way, still significantly larger than implied from the observations. The extremely low value of the $IRX$ ratio may therefore be due to a deficiency in the abundance of the dust, compared to interstellar values, which may reflect the low condensation efficiency of the dust in the circumstellar envelope.
 
Second, the dust deficiency could be the result of grain destruction by thermal sputtering in the hot gas. The sputtering lifetime, $\tau_{sput}$, in a plasma with temperatures above $\sim 10^6$~K is about \citep{Dwe96}:
\begin{equation}
\tau_{sput} \approx 3\times 10^5\ {a(\mu{\rm m}) \over n({\rm cm}^{-3})} \  {\rm yr}
\end{equation}
\noindent
where $n$ is the density of nucleons in the gas.
The X-ray emitting gas is highly ionized, and we will assume that its density is that required to heat the dust to the observed range of temperatures, that is, $n \approx $ 300 to 1400~cm$^{-3}$. Grain destruction is important when the sputtering lifetime is about equal to the age of the shocked gas, which we take to be $\sim$1~yr. The low $IRX$ ratio can therefore be attributed to the effects of grain destruction if the dust particles had initial radii between 10 and 50 \AA. 
So attributing the small $IRX$ ratio to the effect of grain destruction in the hot plasma requires that only small grains had formed in the presupernova phase of the evolution of the progenitor star.

The low $IRX$ ratio shows that IR emission from collisionally heated dust is not the dominant coolant of the shocked gas. Its lower than expected value suggests a dust-to-gas mass ratio in the ER that is only a few percent of its interstellar value. The puzzle of the low dust abundance is greater if, in fact, the IR emission arises from dust that is {\it not} embedded in the X-ray emitting gas, but from dust that resides in the UV/optical knots instead.

\subsection{The Dust Heating mechanism}

The possibility that the initial size of the dust grains swept up by the shock is small suggests that the temperature of the grains may not be at the equilibrium value but may fluctuate due to the stochastic nature of the heating and cooling. Very small grains will be stochastically heated if the energy deposited in the grain in a single collision is large compared to its internal energy content, and if its cooling time via IR emission is shorter than the time between subsequent collisions.  Assuming that the electron and ion temperature is instantaneously equilibrated behind the shock,  the mean thermal energy of the electrons will be 2.6~keV at a postshock temperature of $2\times10^7$~K. Using the electron ranges given by \citet{Isk83} we get that the energy, $\Delta E$, deposited in a dust grain of radius $a(\mu{\rm m})$ and density $\rho$(g~cm$^{-3}$) by electrons with energies $E\gtrsim 370$~eV is given by:
\begin{equation}
\Delta E ({\rm erg}) = 5.5\times10^{-7}\ {\rho\ a\over E({\rm eV})^{0.492}}
\end{equation}
which for $\rho = 3$~g~cm$^{-3}$, $a = 0.0050~\mu$m, and $E=2.6$~keV gives an energy deposition $\Delta E = 1.7\times10^{-10}$~erg. This value is small compared to the internal energy of the grain, which is about $6\times10^{-10}$~erg at the equilibrium grain temperature of 170~K. The average thermal speed of the electrons is about $3\times10^9$~cm~s$^{-1}$, giving an average time between electronic collisions of about 1~sec for an electron density of 300~cm$^{-3}$ and a grain radius of 0.0050~$\mu$m. This is shorter than the grain's cooling time at 170~K, which is about 10~s \citep{Dwe86}, suggesting that the dust can maintain an equilibrium temperature at that value.

Collisions with protons can deposit a significantly larger amount of energy in the dust. Typical proton energies in the shocked gas are $\sim 5$~keV, and the stopping  power for protons with that energy is about 240~MeV~cm$^2$~g$^{-1}$ (NIST tabulated values: {\it http://physics.nist.gov/cgi-bin/Star}). The energy deposited in a 0.0050~$\mu$m radius dust particle is then about $6\times10^{-10}$~erg, which is about equal to the internal energy of the dust grain at 170~K. So collisions with protons and heavier nuclei can be neglected if they are in thermal equilibrium with the electrons, because of their lower collision rate. However, if the electron and ion temperatures are not equilibrated behind the shock, then the dust  heating rate will be dominated by collisions with the protons. The time between successive collisions will be about 40~s for a proton density of 300~cm$^{-3}$. The grains should therefore cool to a temperature of about 100~K before another collision will take place. However, the data does not support such a broad range of dust temperatures, suggesting that the gas density could be somewhat higher (by a factor of $\sim 3$), in which case the dust will cool only to a temperature of $\sim 140$~K, which may be more consistent with the observations. 

All these scenarios support the idea that the dust temperature in the X-ray emitting gas does not fluctuate wildly about the equilibrium dust temperature, which can still provides strong constraints on the density and the equilibration of the electron and ion temperatures in the postshock gas.

\subsection{Dust in the Dense Knots of the Equatorial Ring}

The UVO light emitting knots discovered with the {\it HST} resemble a string of beads uniformly distributed along the ER. 
Figure~\ref{fig14} and Figure~\ref{fig15} depict the map of the dust optical depth overlayed with contour levels of the HST emission obtained on Dec 15, 2004 (day 6502), the closest  to  our $11.7~\micron$ observations. The data look very similar, but  the IR emission seems to emanate from a somewhat wider region than the optical emission, an effect that cannot be entirely accounted for by the lower resolution of the IR data.

Nevertheless the good correlation between the IR emission maps and the {\it HST} image, suggest that a significant fraction, if not most, of the mid-IR emission may be emanating from the knots. The physical conditions of a particular knot (Spot 1 on the ER)   have been modeled in detail by \citet{Pun02}, from the analysis  of the UV/optical line emission detected by the {\it HST} Space telescope Imaging Spectrograph (STIS). They found that the UV fluxes could be fit with a model consisting of two shocks with velocities of $v_s$ = 135 and 250~km~s$^{-1}$ expanding into preshock densities of $n_0$ = 3.3$\times 10^4$ and  10$^4$~cm$^{-3}$, respectively. The postshock temperatures behind the slow and fast shocks are $T_s = 4.5\times 10^5$, and 1.5$\times 10^6$~K,  with gas cooling rates of $\Lambda(T) \approx 10^{-22}$~erg~cm$^3$~s$^{-1}$. 
 The cooling time of the shocked gas is given by $t_{cool} = kT_s / n\Lambda(T_s) $, and is equal to $\approx$ 0.1 and 2~yr for the slow and fast shock, respectively \citep{Pun02}. The thickness of the shock front is therefore $\sim 4\times 10^{13}$ and 8$\times 10^{14}$~cm for the slow and fast shocks, respectively, both significantly smaller than a typical radius of the knot, which is about $2\times10^{16}$~cm.  So most of the dust in the knot resides in the unshocked gas and is heated by the radiation emitted from the cooling shocked gas. 
 
 The radiative energy density seen by  the dust is approximately given by:
 \begin{eqnarray}
 U_{rad} & \approx & {n^2 \Lambda(T_s) \ \ell_{cool} \over c} \\ \nonumber
   & \approx & n\ k\ T_s\ {v_s\over c} \\ \nonumber
    & \approx & f_c\ n_0\ k\ T_s\ {v_s\over c}
\end{eqnarray}
where $\ell_{cool} = t_{cool}\ v_s$, and $f_c$ is the compression factor of the gas in the postshock region. For $T_s =10^6$~K, $n_0 = 10^4$~cm$^{-3}$, and a shock velocity of 200~km~s$^{-1}$ we get that 
\begin{equation}
U_{rad} \approx 2\times 10^{-9}\times f_c \ \ {\rm erg\ cm}^{-3}
\end{equation}
\citet{Pun02} find that compression factors can be as large as $\approx$ 550, giving radiation densities of $\sim 10^{-6}$~erg~cm$^{-3}$ throughout the knot. The energy density of the local interstellar radiation field is about $3\times 10^{-12}$~erg~cm$^{-3}$. Silicate dust particles immersed in  this field achieve equilibrium dust temperatures of about 15~K 
\citep{Zub04}. The energy density in the knot is therefore higher by a factor of $\sim 3\times 10^5$ than that of the local interstellar radiation field, and the average dust temperature should therefore be higher by a factor of $\sim$ 8.3 for a $\lambda^{-2}$ dust emissivity law ($T_d \propto U_{rad}^{1/6}$).  This gives a typical dust temperature of $\sim$ 125~K, in reasonable agreement with the observed average.

The total mass of radiating dust was found to be $\sim 10^{-6}\ M_{\odot}$. The typical mass of gas in a knot of radius $r = 2\times 10^{16}$~cm, and density $n_0 =10^4$~cm$^{-3}$ is $\sim  10^{-4}\ M_{\odot}$. If ${\cal N} \approx$~20 is the number of knots in the ER, then the dust-to-gas mass ratio is $\sim 10^{-6}/({\cal N} \times 10^{-4}) \approx 5\times 10^{-4}$,  or approximately a factor of 10 less than the average dust-to-gas mass ratio in the local interstellar medium. 

The low abundance of dust in the knots could be explained if the dust is efficiently destroyed in the shocked gas {\it and} if  the transmitted shocks have already traversed most of the volume of the knots.  Calculations presented by \citet{Jon04} show that about 49\% of the silicate grains swept up by 200~km~s$^{-1}$ shocks expanding into a medium with a preshocked density of 0.25~cm$^{-3}$ are destroyed. This fraction could be significantly higher for the densities encountered by the shocks traversing the knots. This scenario predicts that the IR emission from the knots was higher in the past, contrary to observed IR light curves (see below). Therefore, if the IR emission emanates from the knots, the low dust abundance must reflect the initial dust abundance in these objects.

\section{THE LIGHT CURVES} 

The light curves at 10 and 20~$\micron$ are shown in Figure~\ref{fig16}: the absolute flux
calibration have been made using \citet{Coh92} 0-magnitude fluxes [$F_0$(10.0~\mic) = 35.24~mJy, 
$F_0$(11.7 \mic) = 28.57~mJy, and $F_0$(18.30~\mic) = 10.25~mJy]. It can be seen from 
Figure~\ref{fig16} that the flux arising from the ejecta at 10~\mic\  declines
exponentially from day 2200 through day 4200 (the ISOCAM and OSCIR
observations) until day 6000, at a rate of $\sim0.32~mag~y^{-1}$.
This would imply that the observations on day 4200 are not dominated by the
ring emission. It is likely that the ring emission
started around day 4000, at roughly the epoch when the first optical spot was discovered 
\citep{Pun97}, and in good agreement with Fig.~5 of \citet{Par02} which presents 
ATCA and {\it Chandra/ROSAT} data. The significant increase of the fluxes reported in the present 
paper compared to the previous fluxes (Fig.~\ref{fig16}) is
consistent with the soft X-ray flux increase observed in the last set of data \citep{Par05b}.
This clearly shows that ``something" has happened at around day $\sim6000$.

This is also manifest in the radio light curve at 843Mhz from MOST. The earlier observations by \citep{Bal01}from 3000 to 6000 days show a steadily increasing flux together with a steadily increasing rate of change. Near day 6000 however \citep{Hun06} the rate of increase undergoes a more abrupt change upwards, and is therefore clearly associated with the changes seen at other wavelengths. 
Figure~\ref{fig16} also shows the Spitzer data point resulting from integrating the IRS data 
within the appropriate bands. Including 
these data points shows that the trend of the increase of the flux at 11.7~\mic\ since 
after day $\sim$6000 
is better fit by a ``linear" function, while the increase in magnitude vs. time is 
not linear. Although we do not make strong conclusions, this could indicate that
the shock is travelling through structures with cross sections (as seen from the center of
the explosion) that do not increase with radial distance, like short cylindrical clouds. 
Sputtering time scales for all but the smallest grains are too long (10--100 years) to affect the IR emission, and the plasma cooling time
due to the IR emission is also too long ($\sim100$ years) to affect the time variation
in the IR emission. 

The 13~cm radio emission light curve (http://www.atnf.csiro.au/research/SN1987A/) is also
shown in Figure~\ref{fig16}. No data are currently available after day 6244 so it is premature to discuss the associated 13 cm evolution. It would be surprising if the rate of increase does not change upwards.
While the sudden increase in the X-ray light curve at day around 3700 has been interpreted
as the encounter of the shock front with the first protrusions of the ER \citep{Par04}, 
the ``jump" after day around 6000 could be the sign of the shock reaching the main body of 
the ER \citep{Par05b}. These authors also show that the light curve of the hard X-rays (3--10 keV) 
is much flatter than one corresponding to the soft X-rays, and similar to the radio light curves, and they argue that 
it is likely that the hard X-ray emission comes from the fast reverse shock rather than the
decelerating forward shock, just like the radio emission \citep{Man05}. 
We note that the reverse shock origin for the radio emission is one possible interpretation for 
explaining the inconsistency of the radio images with the IR and the soft X-rays images.
It is not our intention to discuss evolution of radio fluxes and their relationship to other changes reported in this paper, but we note that the log-log plot in Figure 16 hides a significant change in the rate of increase of the 13 cm at 5000 days. The epoch of increase in the rate appears to be frequency dependent.

\subsection{The Ejecta} 

An asymmetry in the profiles of optical emission lines that appeared at day 530
showed that dust had condensed in the metal-rich ejecta of the supernova \citep{Luc91}.
Although it was discovered via spectroscopy, 
the presence of the dust could be easily inferred from the spectral energy distribution:
as the dust thermalized the energy output, after day 1000 SN~1987A radiated mainly in
the mid-infrared \citep{Bou91}. Although the presence of the dust emission from the
condensates 
in the ejecta was reported in Paper I, we do not detect them in the present observations.
The possibility that this was due to the fact that the present observations were
achieved with the narrow Si5 filter instead of the broad N filter was investigated, 
for the occurence of the [Ne II] $12.8~\micron$ line. Indeed, this line corresponds to a fine structure
transition in the ground state of Ne II so the temperature to excite the upper level
does not need to be high if there is Ne II. However, Paper I reports a dust temperature of 
$90~K < T < 100~K$ in the ejecta, and there is little evidence for X-ray emission from the ejecta. Furthermore, the X-ray emitting gas would be ionized to a much greater extent than to produce NeII. 
It is thus most likely that Ne II is coming from warmer regions near the X-rays which are also responsible for the other HST and {\it Spitzer} lines. 
Why then is the ejecta not detected in our last observations? 

In Paper I, the N band background sigma was 0.033 mJy/pix.
We used a 12 pixels area to integrate the central source, and then the 3-$\sigma$ 
detection limit was 0.34 mJy which is about the 0.32 mJy measurement reported for the 
central source in Paper I.
In the present data, although the source is observed with a better signal-to-noise ratio,
the background at $11.7~\micron$ is affected by a higher noise, with a 3-$\sigma$ background
value of 0.47 mJy. It transpires that we could not have detected a source at the
flux level expected from the radioactive decay based light curve (eg. $\sim0.3$ mJy).
It appears then that the only detection of the ejecta at this late stage is achieved in the
H band (see Figure~\ref{fig7}). The origin of this emission is most likely due to line emission, although we note
that the limiting flux
for any continuum emitter at the center of SN~1987A, in the wavelength range 2900--9650 \AA\  
reported recently by \citet{Gra05} (4.3 mJy in the I filter) is much above our detection at 
H (0.11 mJy). Therefore, our near IR results are compatible with HST observations and do not
exclude contribution from a continuum.

Supernovae are known contributors to interstellar dust. The presence of isotopic anomalies in meterorites \citep{Cla04} and observations of Cas~A \citep{Dou99, Are99} and SNR~1987A (Paper~I and references therein) provide direct evidence for the formation of dust in SNe. However, the relative importance of SNe compared to quiescent outflows from AGB stars in  the production of interstellar dust grains is still unclear \citep{Jon04, Dwe98, Tie98}. So far, the  dust masses in the the ejecta have been determined observationally for two SNe, SN1987A \citep{Luc89} and SN1999em \citep{Elm03}. These masses were 
$\sim~3.10^{-4}$~\Msun\ and $\sim 10^{-4}$~\Msun\ . In both cases the authors note that these values could be much higher if the dust exists in opaque clumps which may be the case. At face value however they are much less than the 
0.1 - 1.0~\Msun\ required to make SNe the dominant source of interstellar dust particles. In future one must find a means of quantifying the effects of clumping. Observations of the handful of other supernovae in which an IR excess is interpreted as dust forming in the ejecta (SN~1979C and SN~1985L, and probably SN~1980K) do not
allow an estimate of the mass of dust. As for the young Galactic SNR which have been observed by IRAS and ISO (Cas~A, Kepler and Tycho) the dust mass deduced is only $10^{-7} - 10^{-3}$~\Msun\ \citep{Lag96, Dwe87c}, also many orders of magnitude lower than the solar mass quantities predicted. 

Paper I discusses the role of supernovae in dust production. We showed there that the dust which condensed in the ejecta of SN~1987A has
survived 16 years since outburst, and was still radiating the energy 
released by the radioactive decay of $^{44}$Ti at the expected level. 
Unfortunately, we could not accurately estimate the mass of the dust, 
and the observations reported in this paper do not allow it either, in the absence 
of detection of the ejecta. 
Thus,
if we consider SN~1987A as an archetype, our data can 
neither support nor rule out the hypothesis that supernovae are significant 
sources for dust production.

\section{CONCLUSIONS}

\begin{figure}
\plotone{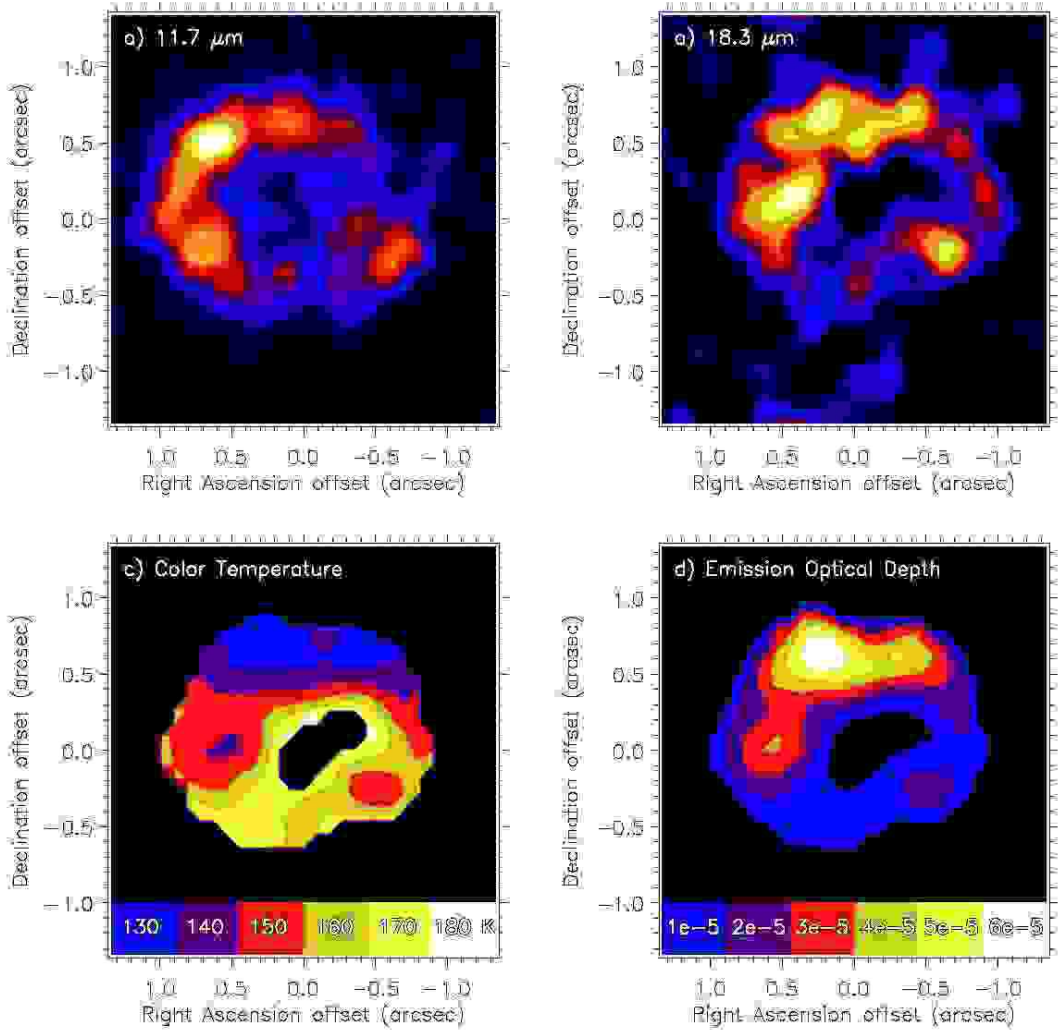}
\caption{(a) SN 1987A seen with T-ReCS at day 6526 in the Si5 narrow band filter 
($11.7~\micron$) and (b) at day 6552 in the Qa filter ($18.3~\micron$); (c) temperature map assuming pure black body emission and \citet{Mat90} extinction law; (d) Opacity map resulting from the same algorithm. These images are smoothed 2 pixels (0.18 arcsec).\label{fig1}}
\end{figure}

\begin{figure}
\plotone{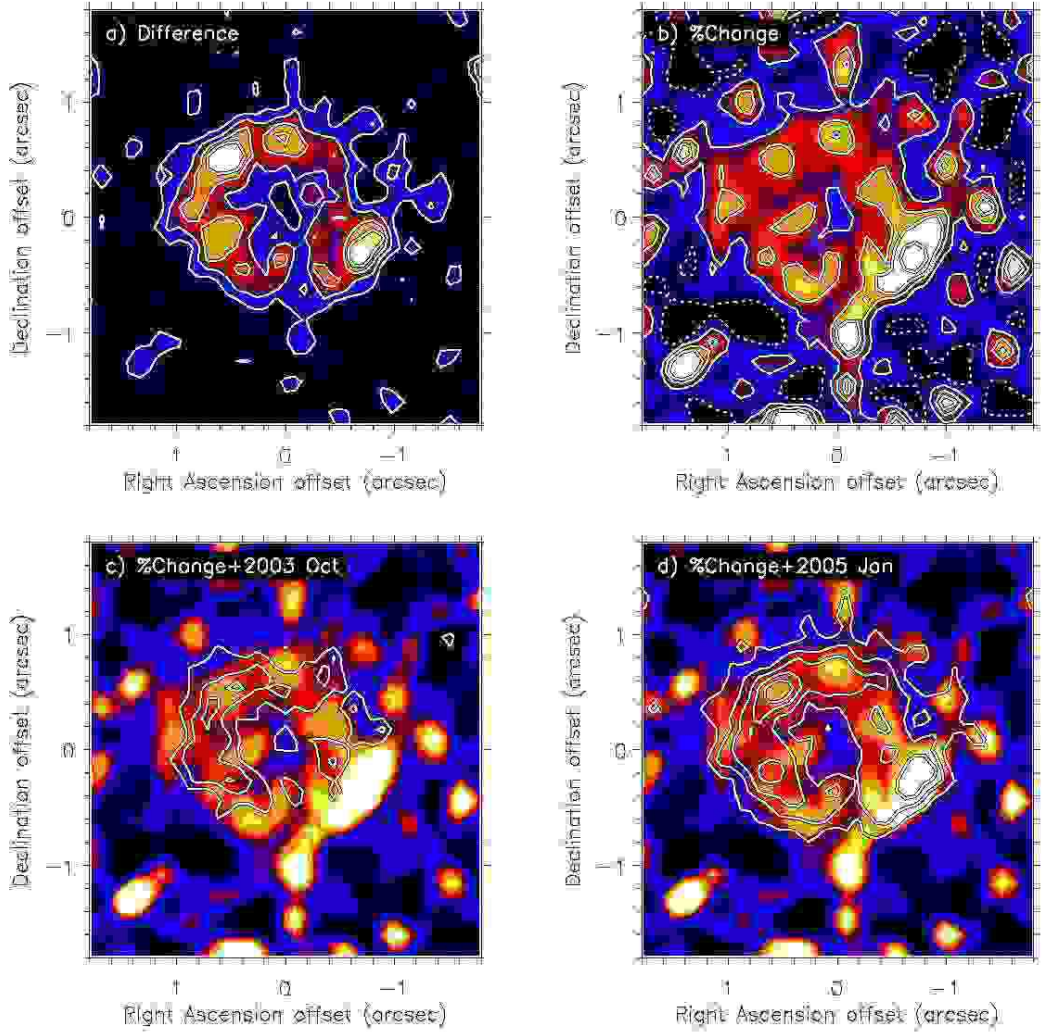}
\caption{ Comparison of the N filter (10~\mic) emission at day 6067 with the
Si5 filter ($11.7~\micron$) emission at day 6526: (a) flux calibrated 
difference of the two images; (b) that difference divided by the 2003 image; contours are drawn at [-50,150,250,350,700] percent changes;(c) image with contours from the 2003 image;(d) image with contours from the 2005 image. All the images are smoothed 2 pixels (0.18 arcsec). \label{fig2}}
\end{figure}

\begin{figure}
\plotone{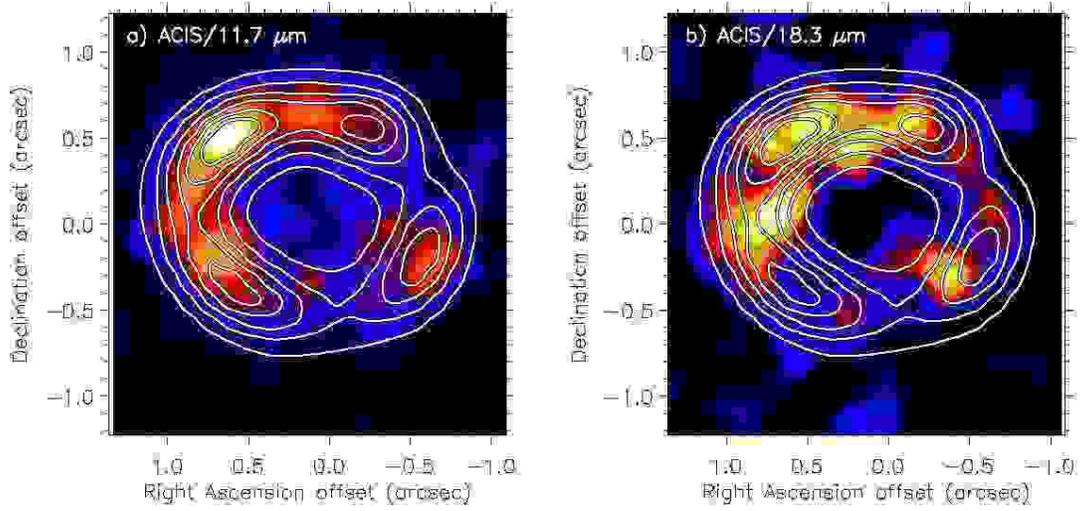}
\caption{Overlays of the contour images obtained with ACIS on Jan. 9-13, 2005
at the {\it Chandra} X-ray Observatory superimposed on the T-ReCS Si5 (a) and Qa (b) image. The T-ReCS image has been smoothed 2 pixels (0.18 arcsec). \label{fig3}} 
\end{figure}

\begin{figure}
\plotone{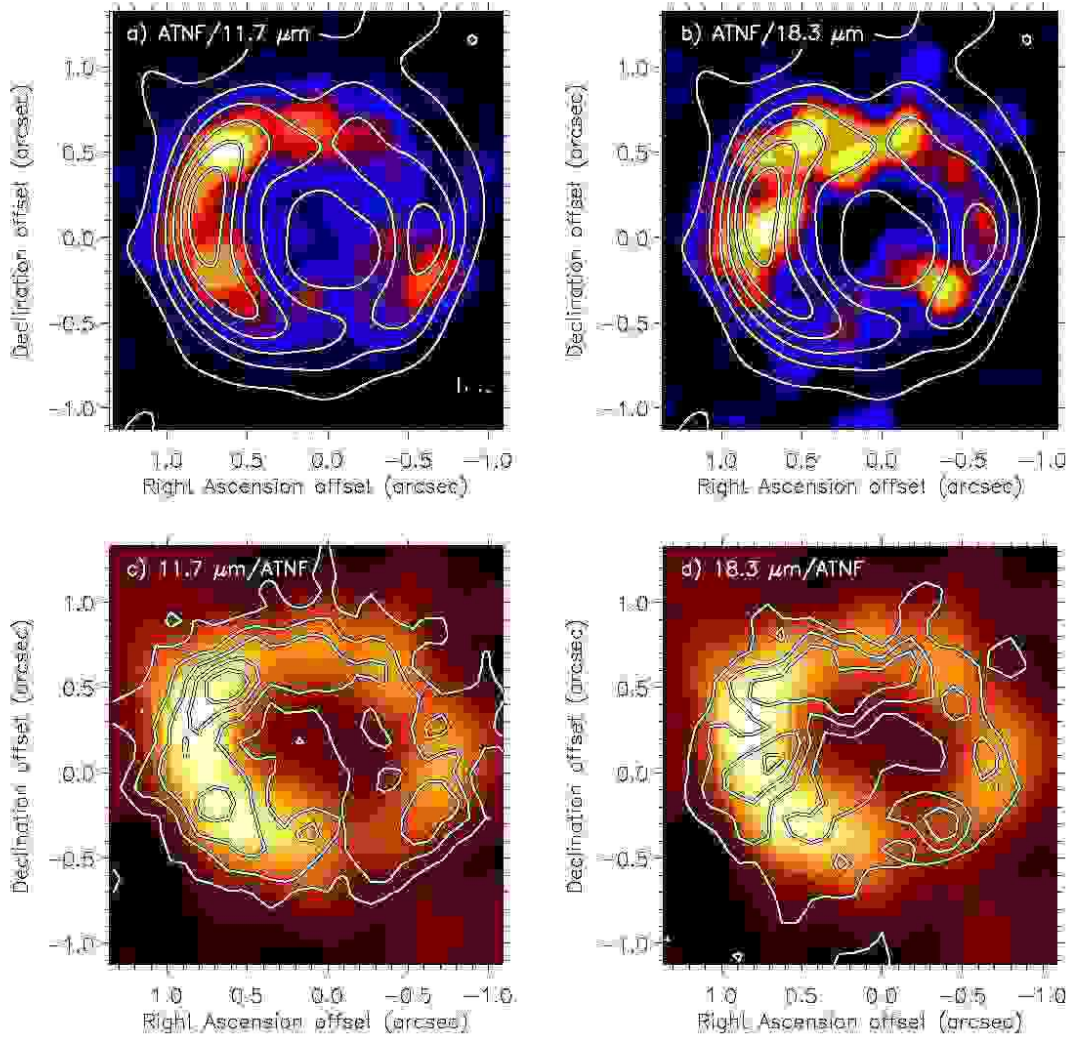}
\caption{ Upper: contours from the ATCA image obtained on Jul. 31, 2003 in the 
12 mm band (16-26 GHz) superimposed on the T-ReCS Si5 (a) and Qa (b) images; lower: ATCA image at same frequencies obtained on May 5, 2004 (ATNF Web page) with contours from T-ReCS at $11.7~\micron$ (c) and $18.3~\micron$ (d). The T-ReCS image have been smoothed 2 pixels (0.18 arcsec).\label{fig4}} 
\end{figure}

\begin{figure}
\plotone{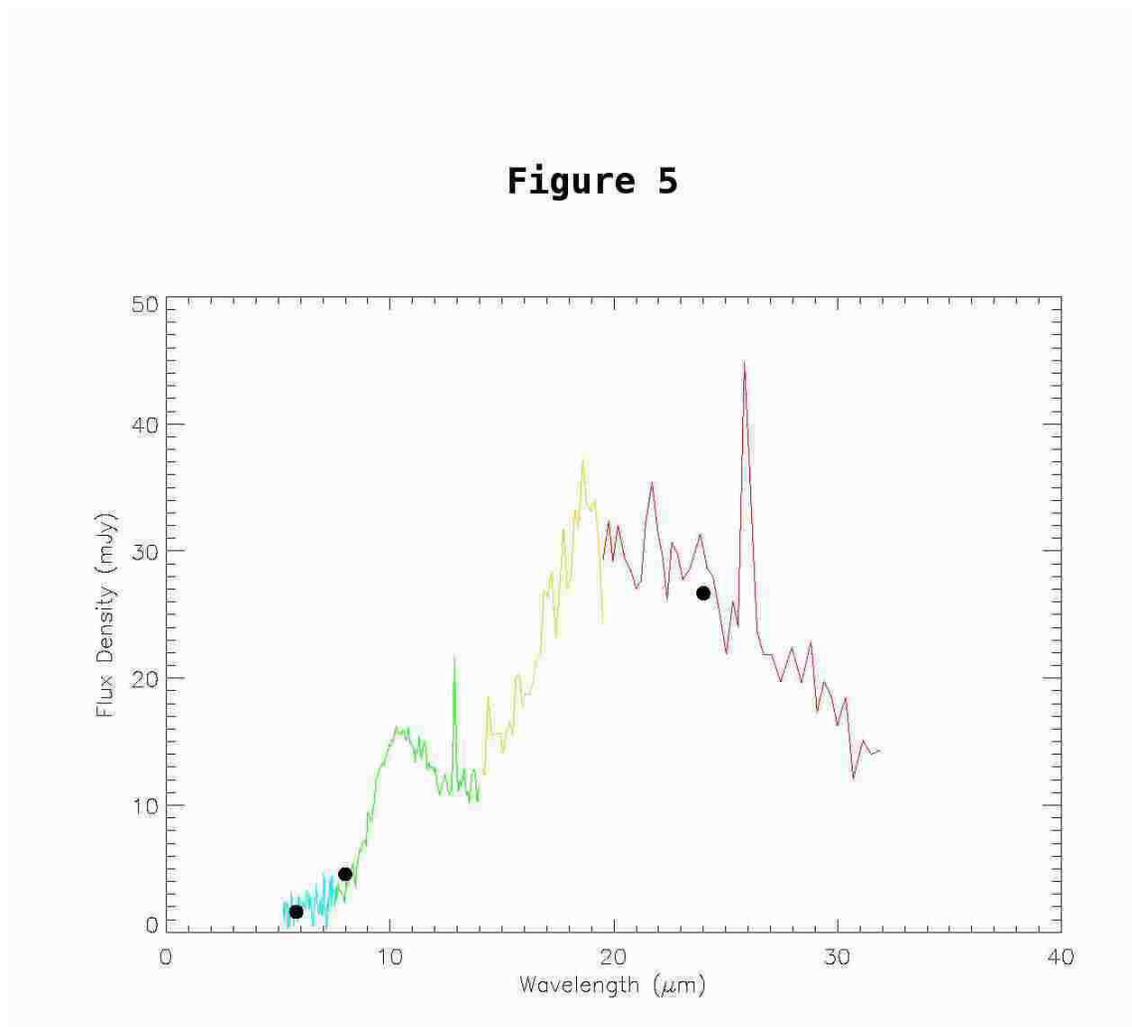}
\caption{The {\it Spitzer} IRS spectrum. The blue and green symbols are from the short low resolution mode; the yellow and red are median binned data from the short and long high resolution modes. The black dots are aperture photometry data points from IRAC and MIPS. \label{fig5}}
\end{figure}

\begin{figure}
\plotone{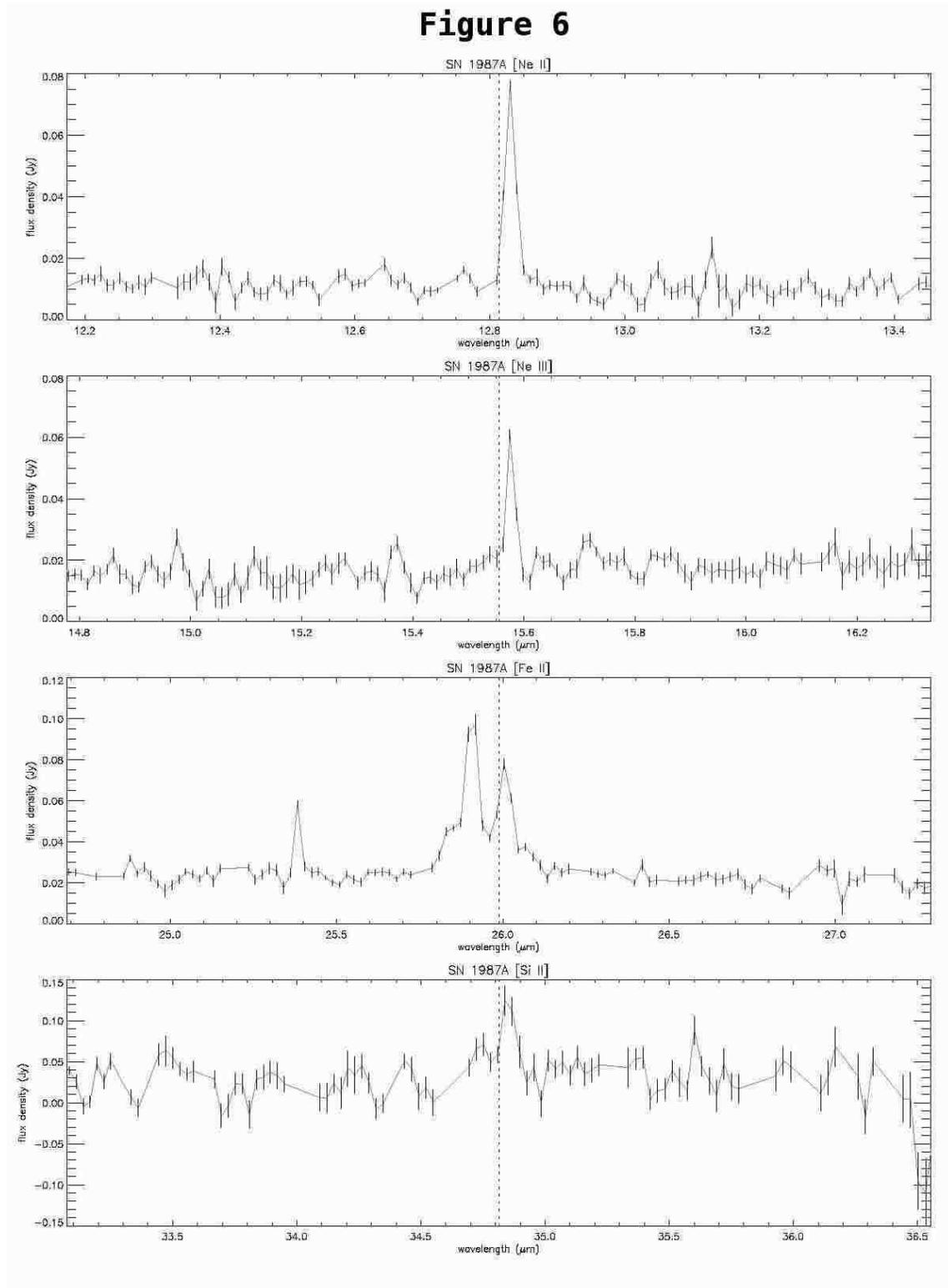}
\caption{Lines emission from SN~1987A observed by the IRS on the{\it Spitzer} Space Telescope (see text). \label{fig6}} 
\end{figure}

\begin{figure}
\plotone{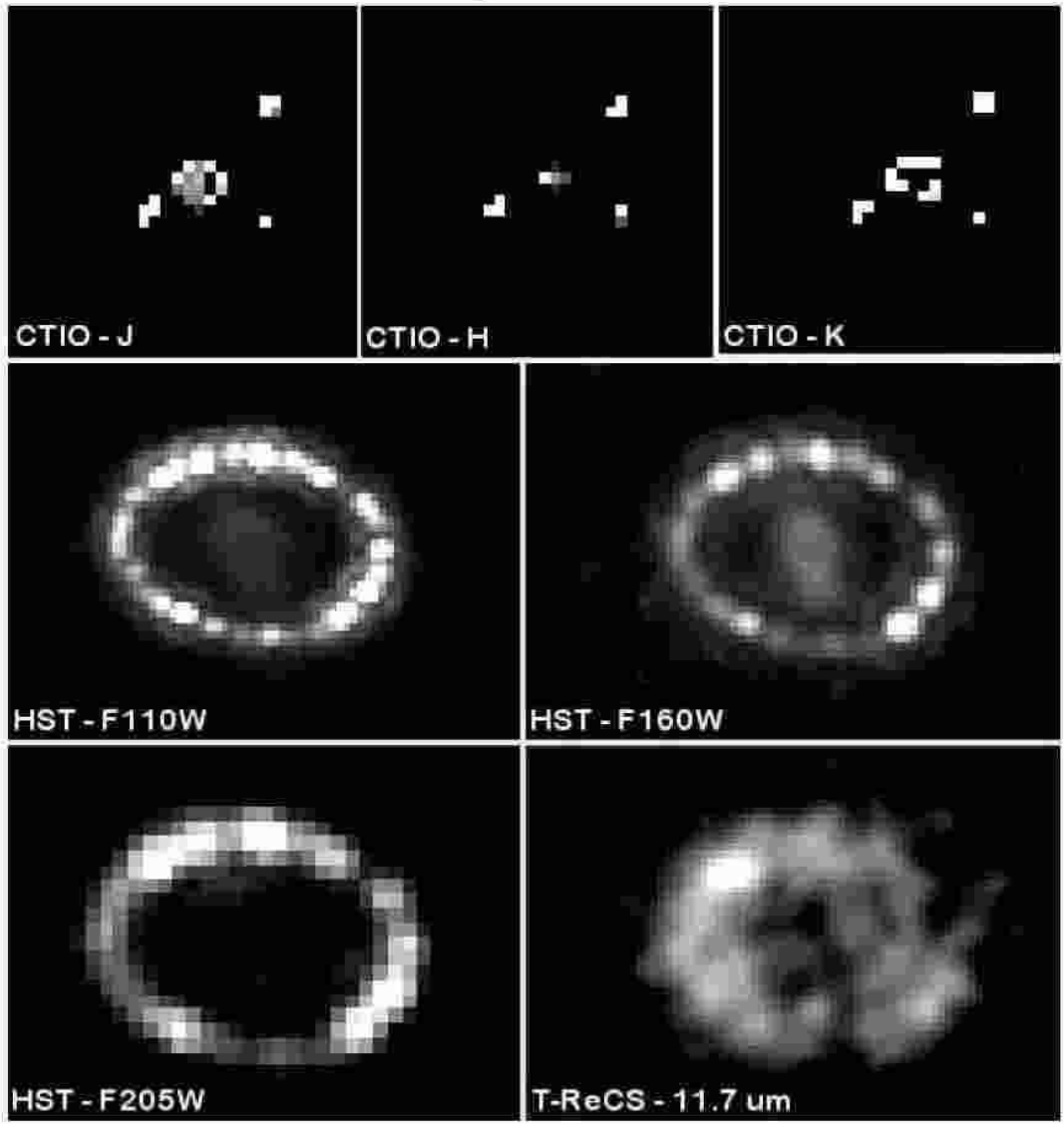} 
\caption{Top: SN~1987A seen with ISPI in the J, H, and K band at the 4-m Blanco telescope at CTIO; the images have been deconvolved using the {\it Multiscale Maximum Entropy Method} from \citet{Pan96}. Bottom: near-IR images obtained with HST on Nov. 11, 2005, in the F110W ($0.8 - 1.4~\micron$), F160W ($1.4 - 1.8~\micron$), and F205W ($1.75 - 2.35~\micron$) filters; the T-ReCS 11.7~\mic\ image is also shown at the same scale. \label{fig7}} 
\end{figure}

\begin{figure}
\plotone{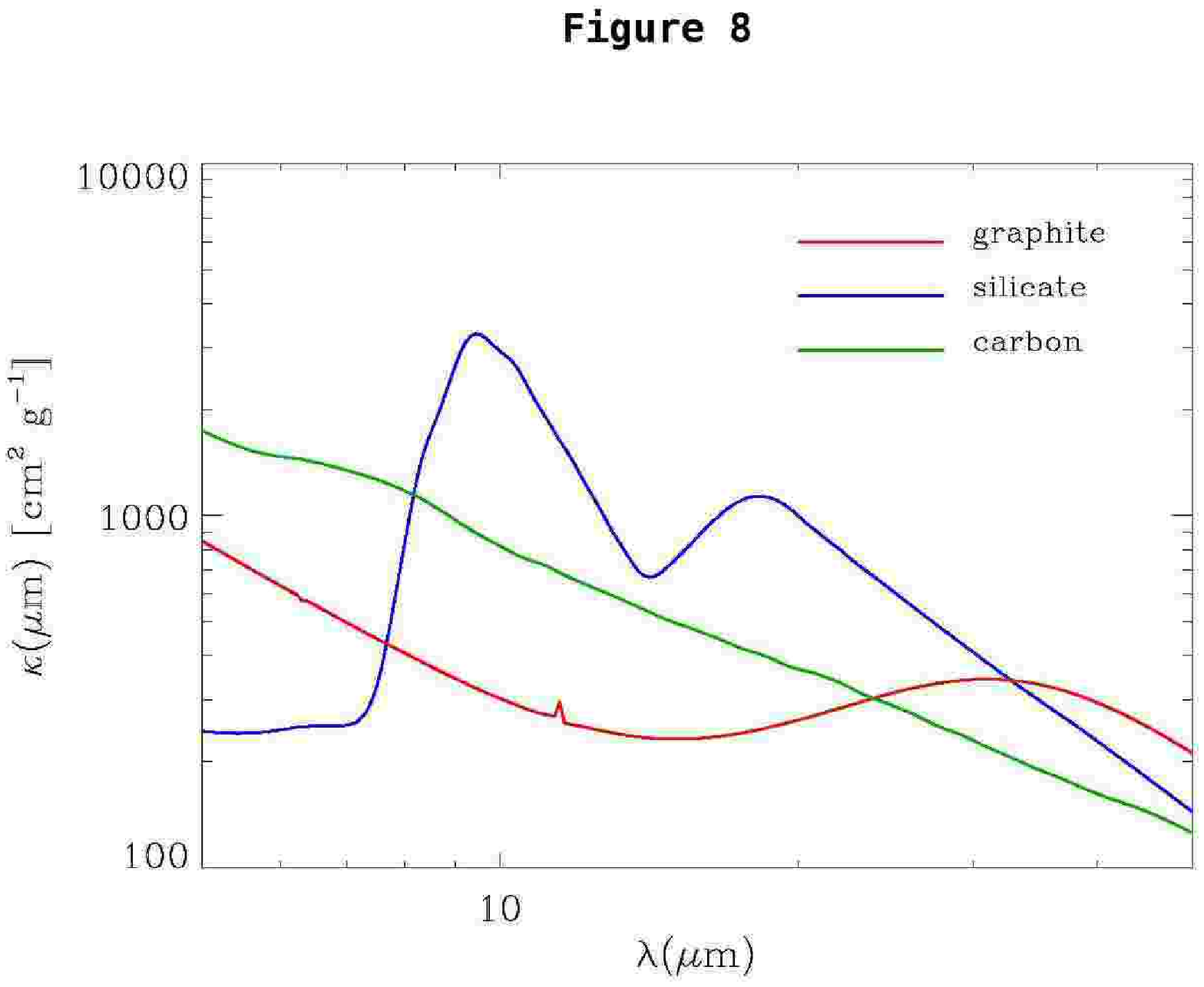}
\caption{The values of the dust mass absorption coefficient, $\kappa(\lambda)$, for the three types of grains considered (references are given in the text). For small  dust particles with radii $\lesssim \lambda$, the value of $\kappa(\lambda)$ is independent of grain radius \label{fig8}. }
\end{figure}

\begin{figure}
\plotone{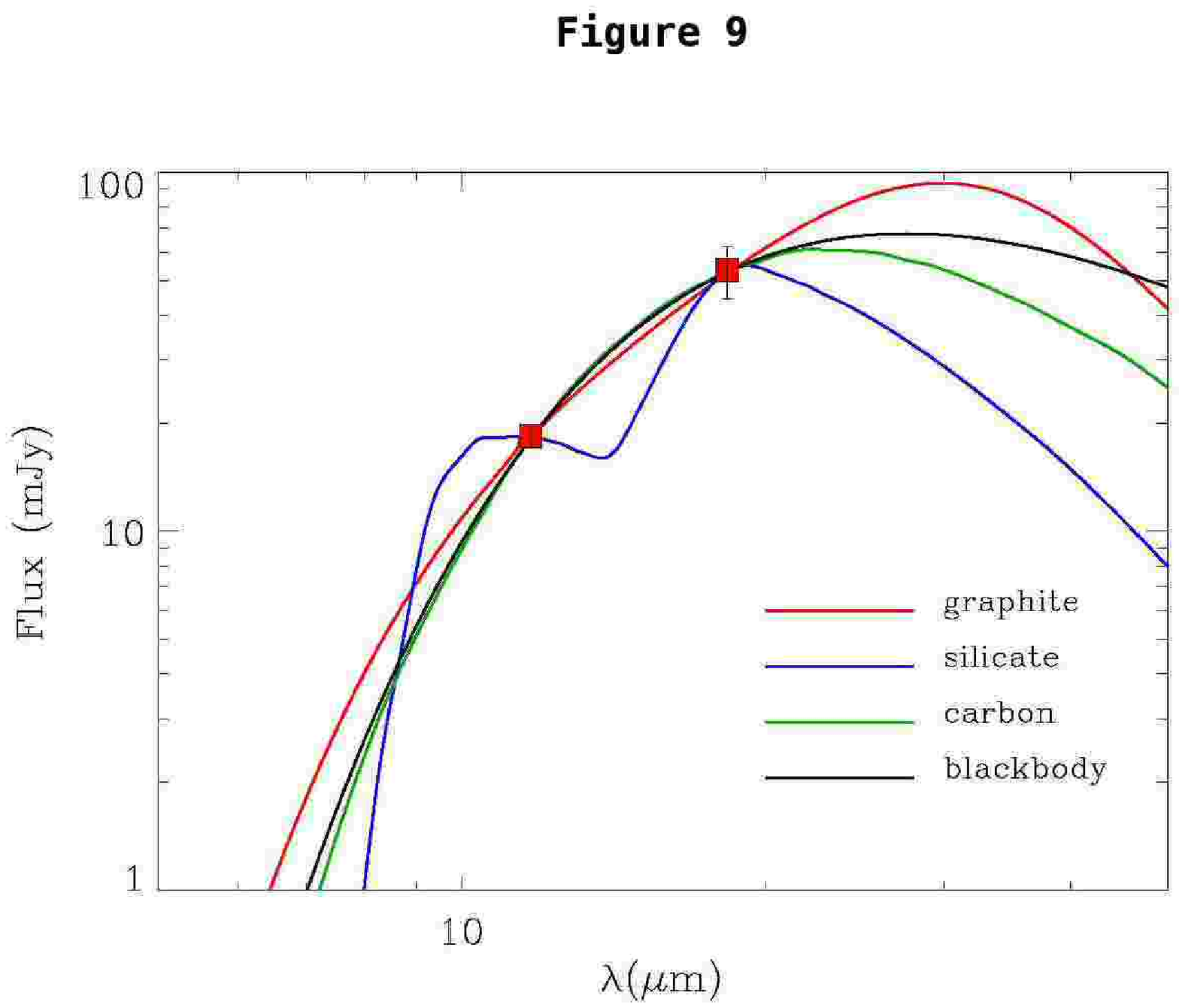}
\caption{ The T-ReCS fluxes at 11.7 and 18.3~$\micron$ fitted by our modelling of the dust continuum for 3 different dust compositions: silicate (blue line), carbon (green line), and graphite (red line) and in the simple black body case (black line). Note that the T-ReCS data alone cannot determine the dust composition. \label{fig9}}
\end{figure}

We have presented mid-IR images of SNR~1987A obtained with T-ReCS on the {\it Gemini} South telescope on day 6526 at 11.7 and 18.3 \mic\ and with IRAC (5.8 and 8~\mic) on day 6130 and MIPS (24~\mic) on day 6187 onboard {\it Spitzer}, together with 3 - 37~\mic\ spectroscopic observations of the remnant obtained with IRS on day 6190 at the same observatory. The imaging observations (Figure~\ref{fig1}) show that the mid-IR emission arises from the  dust in the equatorial ring (ER) heated up by the interaction of the SN blast wave with its circumstellar medium.
Several theoretical models predicted the presence of dust in the CSE
of SN~1987A which was produced in the winds of the supergiant
phase. The location of the IR emission rules out the possibility that the dust condensed out in the SN ejecta, strongly suggesting a circumstellar origin instead. 

\begin{figure}
\plotone{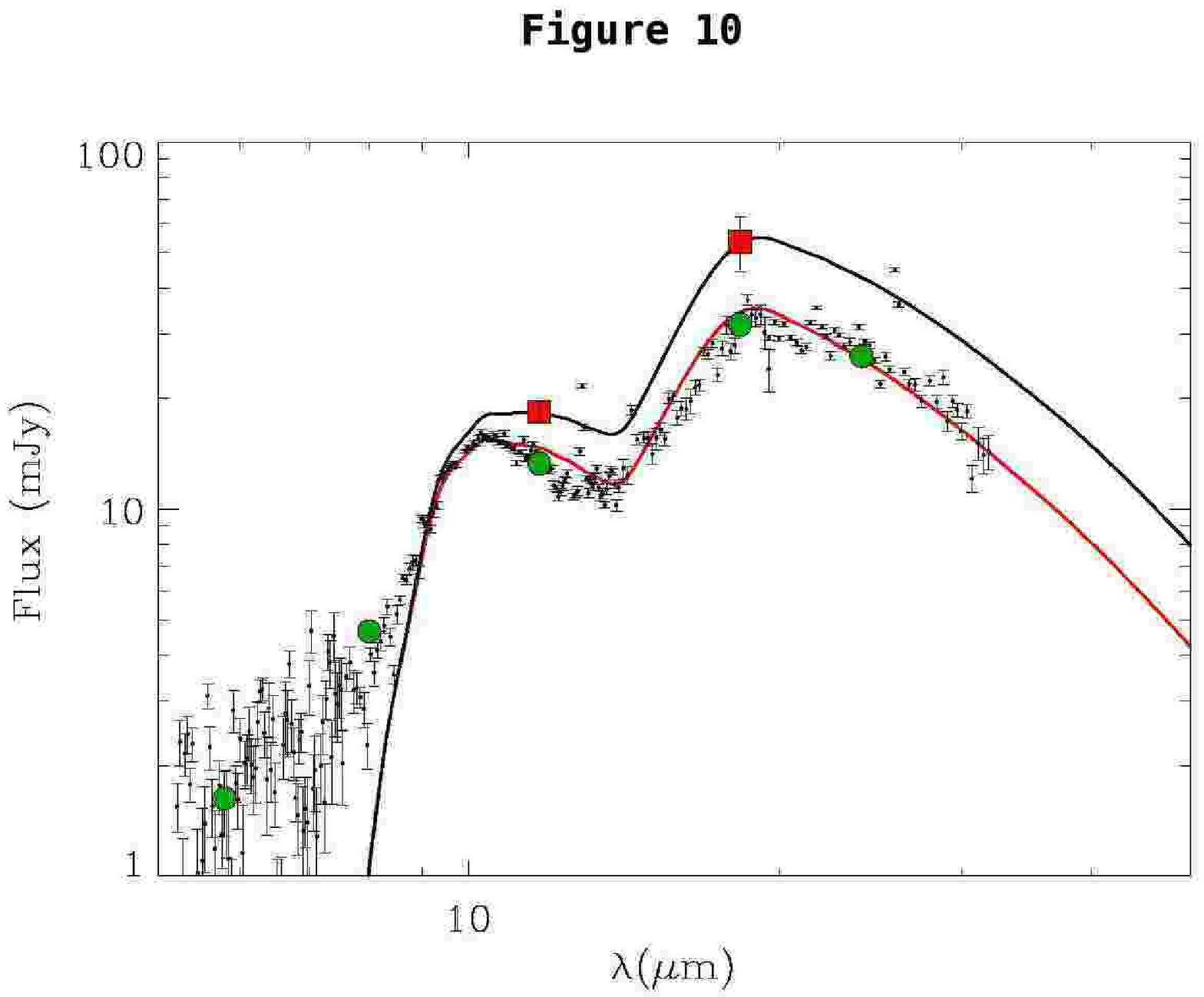}
\caption{The {\it Spitzer} IRS spectrum fitted with a silicate dust model. Green circles:{\it Spitzer} flux measurements. This model has been scaled to fit the T-ReCS flux measurements (red squares). Note that the mid-IR emission has been brightening significantly between both sets of observations. The {\it Spitzer} IRS data unambiguously show that silicate are the major dust component. The derived parameters of the fit are: $T = 180^{+30}_{-20}$~K, $M_{dust} = 0.7-1.7 \times 10^{-6} M_\odot$, and $L_{IR} = 1.6 \pm0.3 \times 10^{36}$ erg s$^{-1}$.\label{fig10}}
\end{figure}

\begin{figure}
\plotone{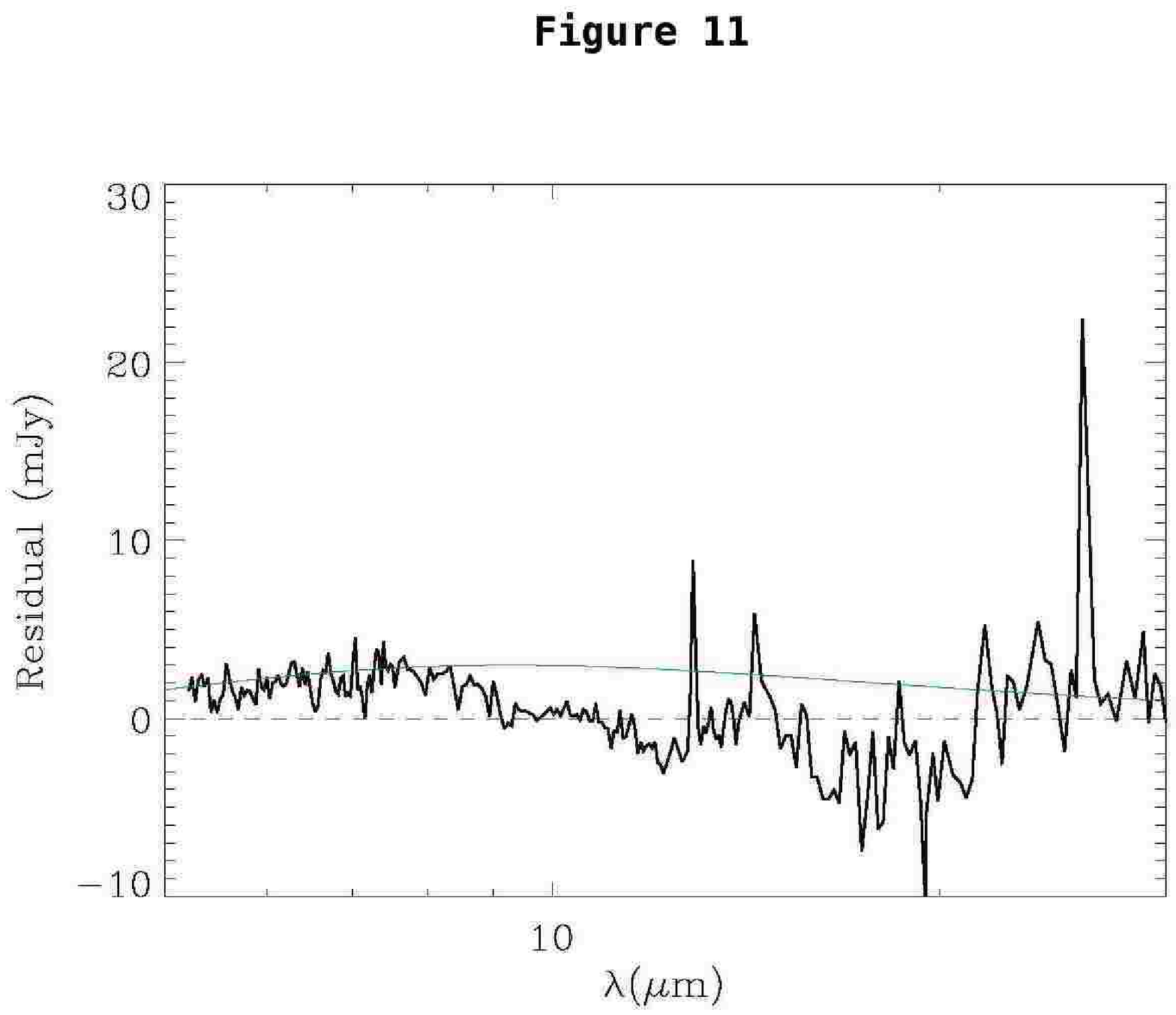}
\caption{Residuals from Spitzer observations fitted with our silicate dust model. The green curve shows emission from amorphous carbon dust at a $T = 750$~K.\label{fig11}}
\end{figure}

\begin{figure}
\plotone{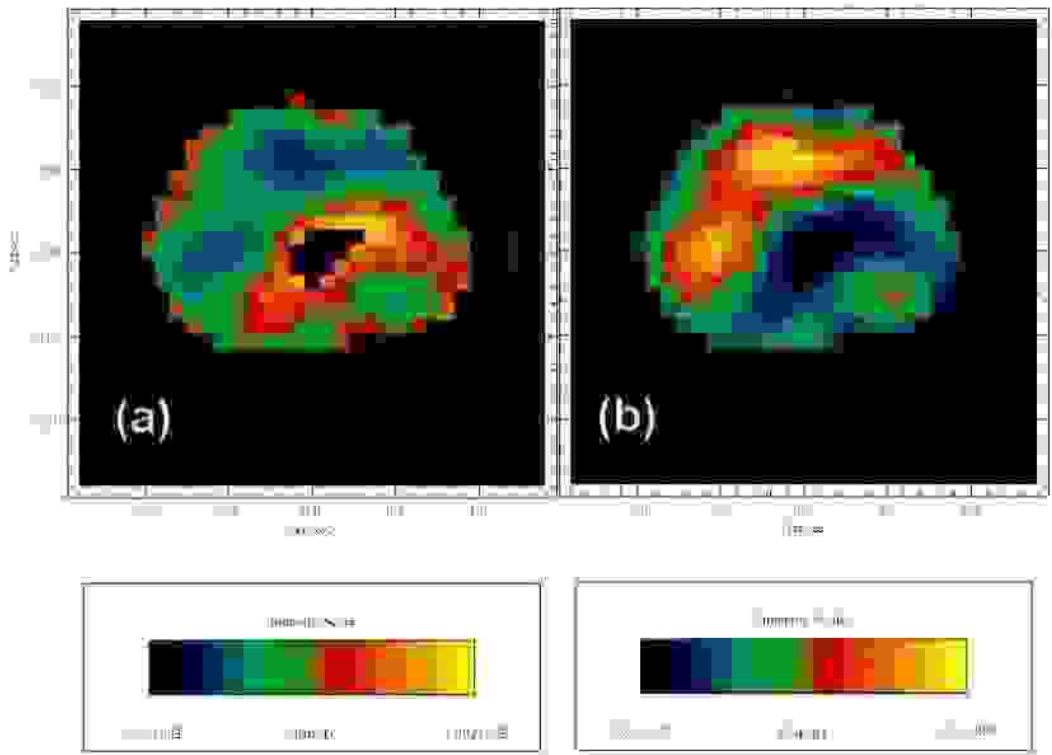}
\caption{ Temperature (a) and optical depth $\tau$(11.7~\mic) (b) maps derived 
for the T-ReCS observations for a silicate dust composition. \label{fig12}} 
\end{figure}

\begin{figure}
\plotone{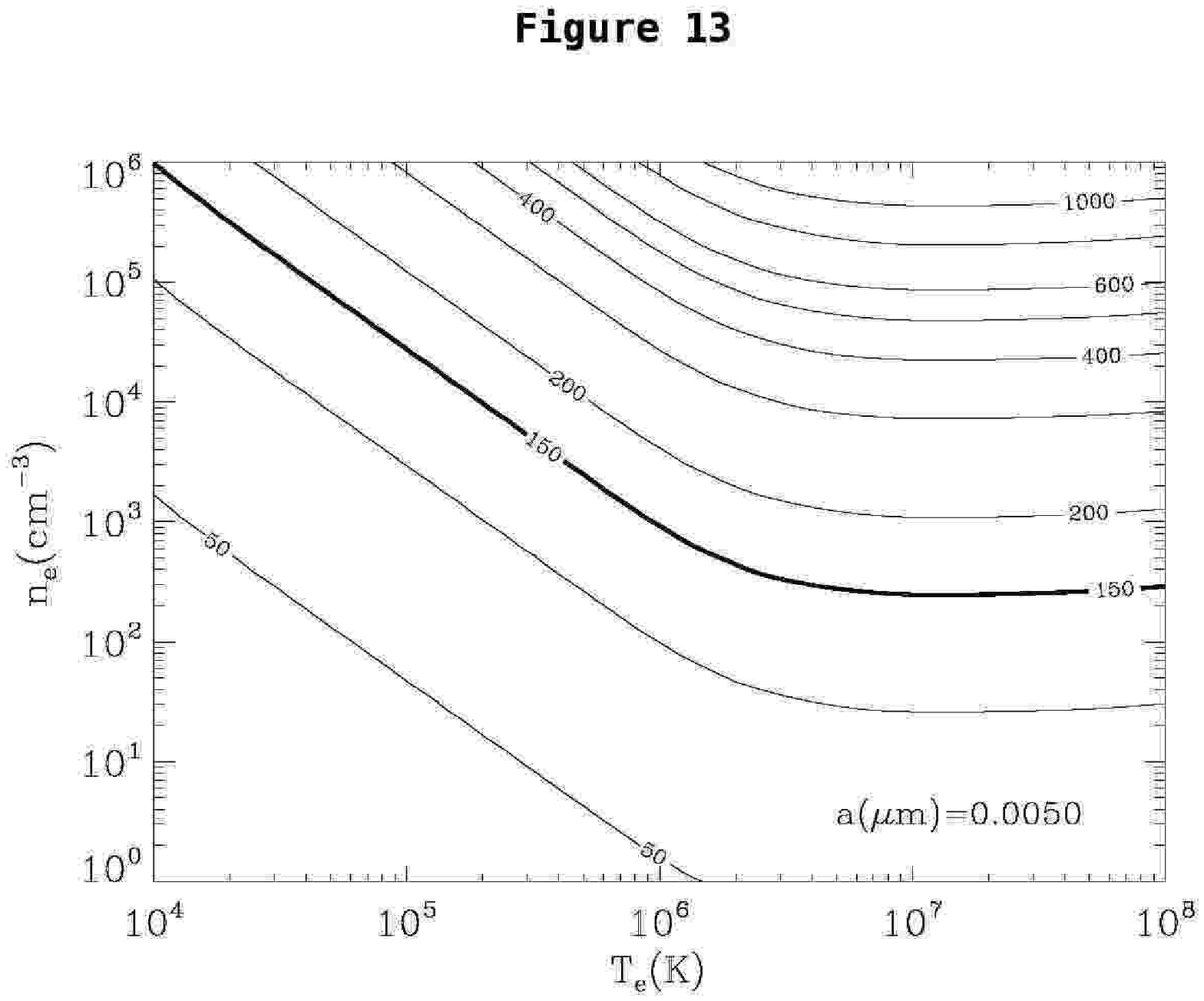} 
\caption{ Contours of silicate dust temperature collisionally heated by a hot gas as a function of plasma temperature and electron density. \label{fig13}} 
\end{figure}

\begin{figure}
\plotone{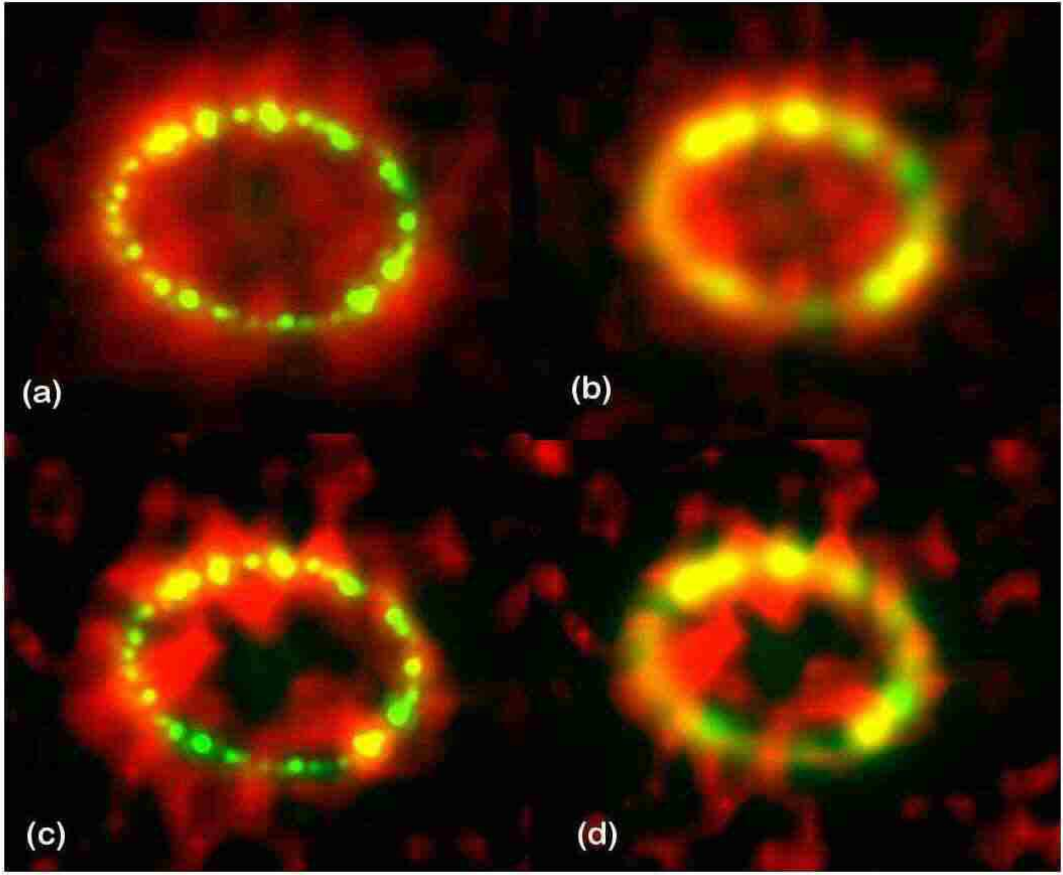}
\caption{(a) Montage of the mid-IR 11.7~\mic\ image of the ER (red) and the HST F625W image obtained on day 6502 (yellow);
(b) same as (a) with the HST image (yellow) convolved to match the resolution of the T-ReCS 11.7~\mic\ image (red); (c) montage of the Qa image from T-ReCS (red) with the HST F625W image (yellow); (d) same as (c) with the HST image convolved to match the T-ReCS Qa angular resolution. \label{fig14}} 
\end{figure}

\begin{figure}
\plotone{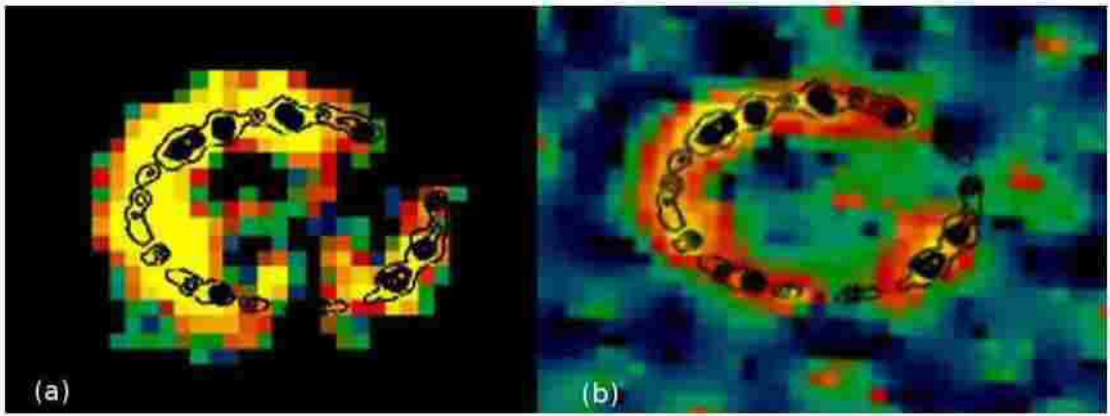}
\caption{(a): Image of the ER obtained through the 11.7~\mic\ filter with T-ReCS with contours of the {\it HST} UVO image obtained on day 6502 superimposed; (b): same with the 11.7~\mic\ image deconvolved with the {\it ``Multiscale Maximum Entropy Method"} from \citet{Pan96}. \label{fig15}}
\end{figure}

\begin{figure}
\plotone{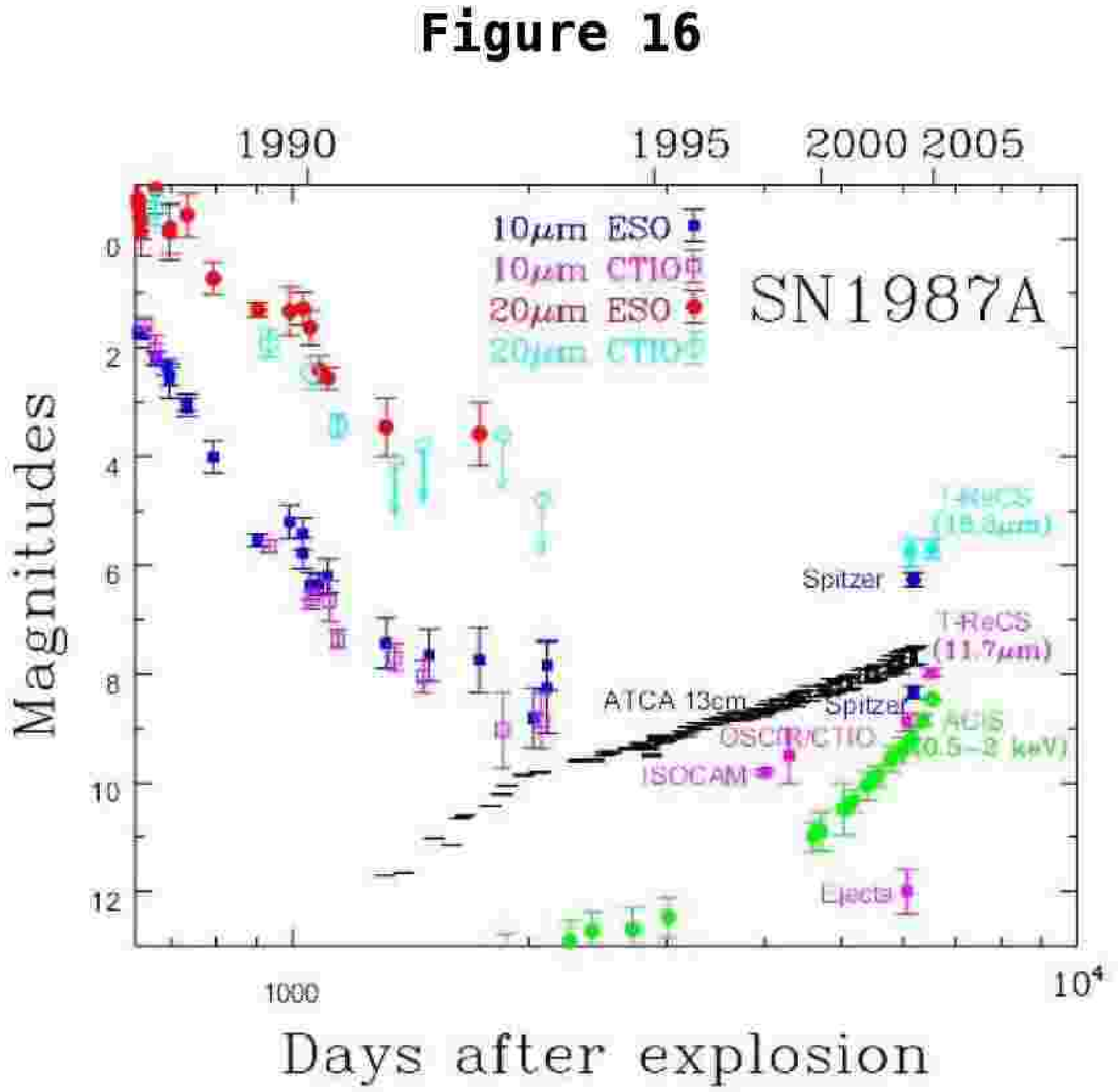}
\caption{ Light curves of SN 1987A for the indicated frequencies; the X-ray data points before day $\sim4000$ are from the
ROSAT data \citep{Has96}; ATCA and ROSAT/ACIS fluxes are plotted in arbitrary logaritmic scales; ATCA data points are from the ATNF web page (http://www.atnf.csiro.au/research/SN1987A/)
and ACIS data are from \citet{Par05b}. The upper point in the {\it Spitzer}/IRS data is integrated in the 18.3~\mic\ bandpass, and the lower point
is integrated in the 11.7~\mic\ bandpass.The point called `Ejecta'
derives from the point source near the center of the image reported in Paper I. After day 4000 all data points refer to the inner ER. \label{fig16}}
\end{figure}

\begin{figure}
\plotone{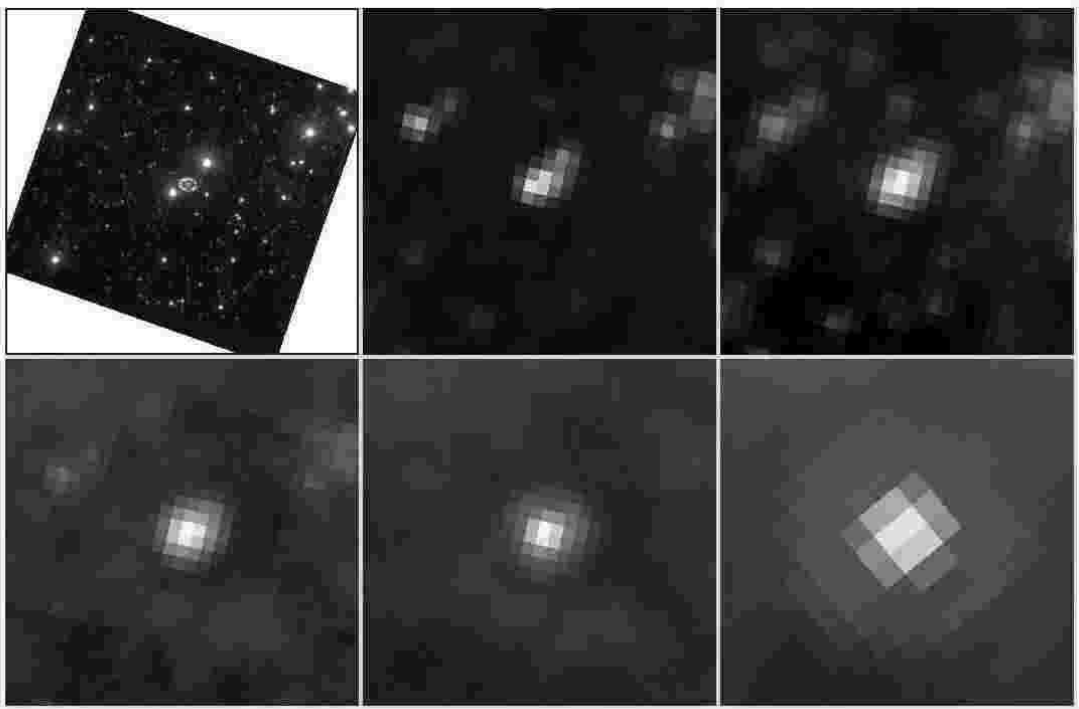}
\caption{Mid--IR images of SN 1987A from Spitzer, compared to a 
near--IR image from the Hubble Space Telescope.
Left to right, top to bottom: (a) Hubble ACS/HRC F814W (range = [30,  
3.2e4] electrons, pixel = $0.05"$);
(b) Spitzer IRAC 3.6 $\mu$m (range=[-0.2, 4] MJy/sr, pixel = $1.2"$);
(c) Spitzer IRAC 4.5 $\mu$m (range=[0.2, 5] MJy/sr, pixel = $1.2"$);
(d) Spitzer IRAC 5.8 $\mu$m (range=[1.8, 8] MJy/sr, pixel = $1.2"$);
(e) Spitzer IRAC 8 $\mu$m (range=[2, 20] MJy/sr, pixel = $1.2"$);
(f) Spitzer MIPS 24 $\mu$m (range=[18, 45] MJy/sr, pixel = $2.45"$).
The bright stars in the Hubble image NW and SE of SN 1987A
(Star 2 and Star 3) appear to be brighter than the SN emission at
3.6 $\mu$m, but rapidly fade as the SN brightens at longer wavelengths.
The large ring in the MIPS image is the first diffraction ring of the
PSF, which is visible for any bright unresolved source.
All pictures are oriented with N upward and E to the left. All span
a $35.2" \times 34.6"$ field.
The Hubble image is a preview image from the Multimission Archive
at Space Telecope (MAST). The Spitzer images are post-BCD images
from the science data archive at the Spitzer Science Center. \label{fig17}}
\end{figure}

\begin{figure}
\plotone{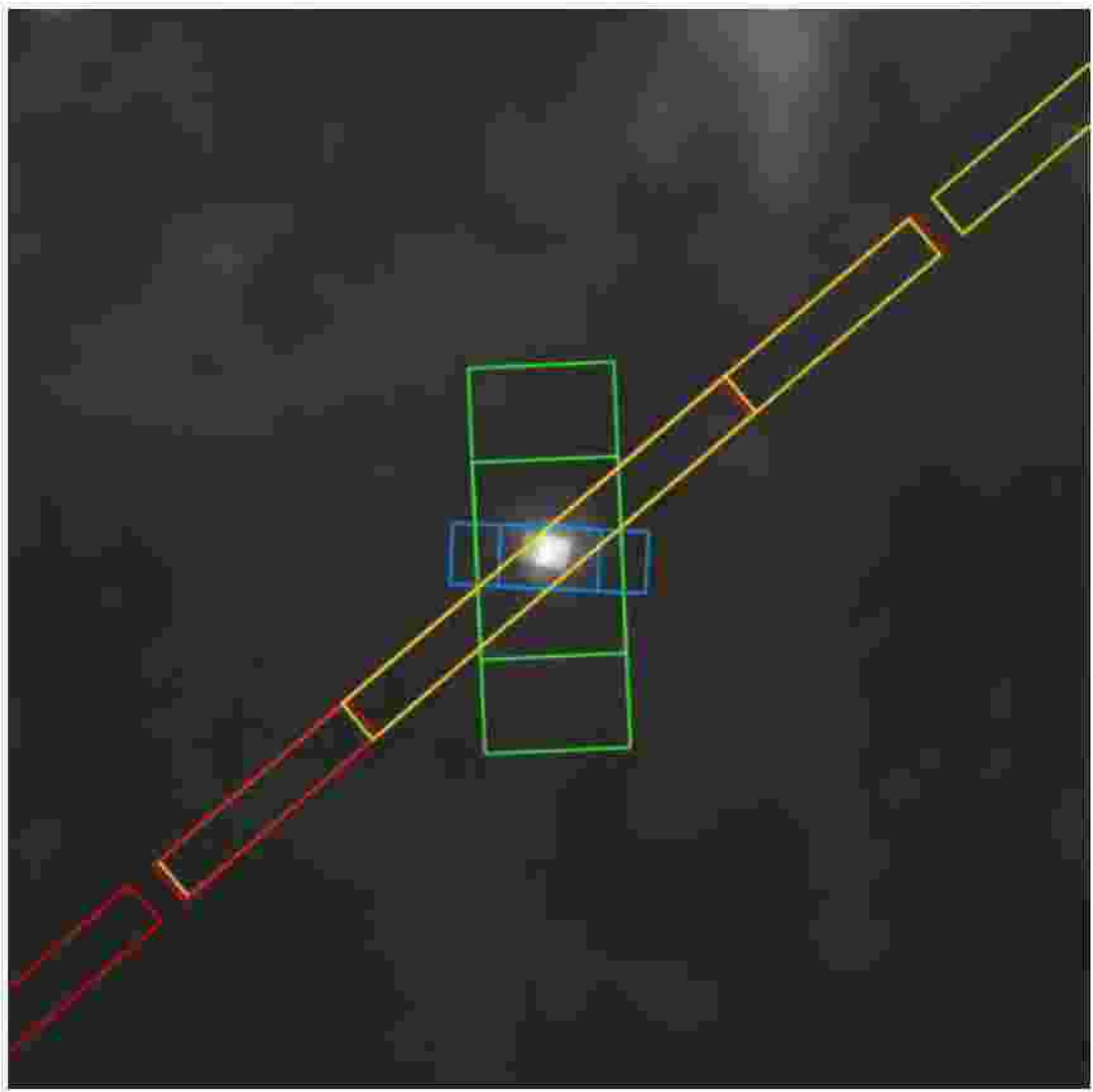}
\caption{IRS slit positions overlaid on the Spitzer IRAC 8 $\mu$m image
(range = [0,15] MJy/sr).
This image spans a $82.8'$ square field. The slit positions
indicated by the long narrow red/yellow rectangles are the
IRS short-lo (SL) 1st/2nd order (7.4-14.2 / 5.3-8.5 $\mu$m) pointings.
The short narrow blue rectangles indicate the short-hi (SH) pointings.
The short wide green rectangles indicate the long-hi (LH) pointings.
For each IRS module the target is positioned at locations 1/3 and 2/3
of the distance along the slit length. \label{fig18}}
\end{figure}

The 3 - 37~$\micron$ spectrum (Figures~\ref{fig10}) shows that the emission arises from a population of astronomical silicate particles. Temperature maps show that the dust temperature is fairly uniform in the ER and about 166$^{+18}_{-12}$~K, with total dust masses of $\sim 2.6^{+2.0}_{-1.4} \times 10^{-6}$~\Msun. 

A comparison with {\it Chandra}  and {\it HST} observations show an equally good correlation between the 11.7~\mic\ IR and the X-ray and  UV-optical images of the supernova. Because of the limited angular resolution in the IR we cannot determine the location or heating mechanism of the radiating dust.

The dust could be residing in the hot $\sim 10^7$~K gas and collisionally heated by the X-ray emitting plasma. The dust temperature is then an excellent diagnostic of the electron density, giving a value of $\sim $ 300-1400~cm$^{-3}$, similar to the value suggested by the {\it Chandra} observations. 

Comparison of the IR and X-ray fluxes suggests that the dust is depleted by a factor of $\sim$ 30 in the X-ray emitting gas,  compared to its value in the local interstellar medium of the LMC.
This low value could be due to its destruction by thermal sputtering in the shocked gas, requiring the initial grain radii to be below $\sim$ 50~\AA. 

Alternatively, the dust could be residing in the UV-optical emitting knots in the ER and radiatively heated by the cooling gas that was excited by the shocks propagating through these knots. A simple calculation for a particular knot shows that the radiative energy density can heat the dust to typical temperatures of about 125~K, similar to those inferred from the IR observations. A comparison with the mass of the knots shows that the dust-to-gas mass ratio in the knots is lower by a factor of $\sim$ 10 compared to  its value in the ISM of the LMC. This low abundance reflects the low condensation efficiency of the dust in the outflow of the progenitor star.

We stress that in order to assess the role of SNe in the production of dust in
the Universe, it is clearly important to measure
the presence of dust that survives into the formation of the remnant,
and for this, mid-IR and sub-mm observations are critical.

\acknowledgements{PB is most grateful to Eric Pantin for the use of his {\it ``Multiscale Maximum Entropy Method"} program for the deconvolution of the images, and for helpful discussions related to it. The authors acknowledge R. Manchester for 
providing the ATCA image obtained on July, 31, 2003. This work is based in part on observations made with the
Spitzer Space Telescope, which is operated by the Jet Propulsion Laboratory, California Institute of Technology, under a contract with NASA. SMART was developed by the IRS Team at Cornell University and is available through the Spitzer Science Center at Caltech.
JD acknowledges support by grants from MIUR COFIN, Italy.
NBS acknowledges support for the study of SN~1987A
though the HST grants GO-8648 and GO-9114 for the Supernova INtensive
Survey (SInS: Robert Kirshner, PI). SP was in parts supported by the SAO grant GO5-6073X. ED was supported in part by NASA LTSA 2003.}

\clearpage

\begin{deluxetable}{lccccccc}
\tablecaption{Observed Fluxes From SN1987A\tablenotemark{1} \label{tab1}}
\tablehead{
\colhead{Day\tablenotemark{2}} &
\colhead{Instrument} &
\colhead{ 5.8 $\mu$m} &
\colhead{ 8.0 $\mu$m} &
\colhead{ 10.4 $\mu$m} &
\colhead{ 11.7 $\mu$m} &
\colhead{ 18.3 $\mu$m} &
\colhead{ 24 $\mu$m}
}
\startdata
6067 &  T-ReCS 				& \nodata 		&  \nodata 		&    9.9$\pm$1.5 & \nodata & \nodata &  \nodata \nl
6125 &  T-ReCS 				& \nodata 		&  \nodata 		& \nodata 		 &  \nodata & $<$ 50.6 &  \nodata \nl
6130 &  IRAC   				& 1.79$\pm$0.06	&  4.98$\pm$0.16 & \nodata 		 &  \nodata & \nodata &  \nodata \nl
6190 &  IRS\tablenotemark{3} & \nodata 		&  \nodata 		& \nodata 		 &  13.3$\pm$0.3 & 31.8$\pm$1.0 &  \nodata \nl
6184 &  MIPS 				& \nodata 		&  \nodata 		& \nodata 		 &  \nodata 	& \nodata &   29.8$\pm$ 1.7 \nl
6487 &  IRAC   				 & 2.44$\pm$0.19 &  7.48$\pm$0.19 & \nodata        &  \nodata & \nodata &  \nodata \nl
6526 &  T-ReCS 				& \nodata 		&  \nodata 		& \nodata 		 &  18.4$\pm$1.2 & 53.4$\pm$9 &  \nodata \nl
\enddata
\tablenotetext{1}{Fluxes in units of mJy.}
\tablenotetext{2}{Time elapsed since the explosion.}
\tablenotetext{3}{{\it Spitzer} IRS data integrated between the filter half-power wavelengths.}
\end{deluxetable}

\begin{deluxetable}{llllll}
\tablecaption{Measured IR Lines from SN1987A at Day 6190 \label{tab2}}
\tablehead{
\colhead{Line Center ($\mu$m)} &
\colhead{Element} &
\colhead{FWHM (km s$^{-1}$)} &
\colhead{Line Flux (W cm$^{-2}$)} &
\colhead{S/N} &
\colhead{Note}
}
\startdata
$12.831 \pm 0.001$ & [Ne II]  & 442 & $2.392 \pm 0.185 10^{-21}$ & 20 & 1 \nl
$14.337 \pm 0.004$ & [Ne V]   & 960 & $7.153 \pm 2.043 10^{-22}$ &  4 & 2 \nl
$15.577 \pm 0.001$ & [Ne III] & 336 & $1.038 \pm 0.141 10^{-21}$ & 13 & 1 \nl
$24.340 \pm 0.002$ & [Ne V]   & 374 & $4.819 \pm 1.265 10^{-22}$ &  9 & 3 \nl
$25.833 \pm 0.004$ & [F IV]   & 472 & $4.613 \pm 1.163 10^{-22}$ &  7 & 4 \nl
$25.904 \pm 0.001$ & [O IV]   & 635 & $2.016 \pm 0.122 10^{-21}$ & 24 & 5 \nl
$26.005 \pm 0.002$ & [Fe II]  & 844 & $1.683 \pm 0.145 10^{-21}$ & 15 & 5 \nl
$34.850 \pm 0.008$ & [Si II]  & 603 & $1.789 \pm 0.616 10^{-21}$ &  4 & 1 \nl
\enddata
\tablenotetext{1}{{Line is strong in background.}}
\tablenotetext{2}{{Line was only recognized by searching for counterpart to the 24.3 $\mu$m line.}}
\tablenotetext{3}{{Line is faintly visible in 2-D spectral data.}}
\tablenotetext{4}{{Line is not well-resolved. It could also be fit as the wing of a very broad line
underneath the [O~IV] and [Fe~II] lines.}}
\tablenotetext{5}{{Line is weak or absent in background.}}
\end{deluxetable}

\begin{deluxetable}{ccccc}
\tablecaption{Global parameters from Silicate Dust Modeling\tablenotemark{1} \label{tab3}}
\tablehead{
\colhead{Instrument} &
\colhead{Day\tablenotemark{2}} &
\colhead{$T_{dust}$ (K)} & 
\colhead{$L_{IR}$ (erg s$^-1$)} &
\colhead{$M_{dust}$ ($M_\odot$)}  
}
\startdata
T-ReCS  & 6070 & $180^{+20}_{-10}$ & $9\pm3 \times 10^{35}$ & $1-4 \times 10^{-6}$ \nl 
Spitzer & 6190 & $180^{+20}_{-15}$ & $1.6\pm0.3 \times 10^{36}$ & $0.7-1.7 \times 10^{-6}$ \nl
T-ReCS  & 6530 & $166^{+18}_{-12}$ & $2.3\pm0.4 \times 10^{36}$ & $1.2-4.6 \times 10^{-6}$  \nl 
\enddata
\tablenotetext{1}{The dust mass for day 6070 has been
recalculated from Paper I with the value of $\kappa$ adapted to the silicate case.} 
\tablenotetext{2}{Time elapsed since the explosion}
\end{deluxetable}

\clearpage


\end{document}